\documentclass[a4paper]{jpconf}
\usepackage{graphicx}
\usepackage{mathrsfs,amsmath,amssymb,amsfonts}
\begin{document}
\title{Pygmies, Giants, and Skins}

\author{J. Piekarewicz}

\address{Department of Physics, Florida State University,
                  Tallahassee, FL 32306-4350, USA}

\ead{jpiekarewicz@fsu.edu}

\begin{abstract}
 Understanding the equation of state (EOS) of neutron-rich matter 
 is a central goal of nuclear physics that cuts across a variety of
 disciplines.  Indeed, the limits of nuclear existence, the collision
 of energetic heavy ions, the structure of neutron stars, and the
 dynamics of core-collapse supernova all depend critically on the
 nuclear-matter EOS. In this contribution I focus on the EOS of cold
 baryonic matter with special emphasis on its impact on the structure,
 dynamics, and composition of neutron stars.  In particular, I discuss 
 how laboratory experiments on neutron skins as well as on Pygmy 
 and Giant resonances can help us elucidate the structure of these
 fascinating objects.
 \end{abstract}

\section{Motivation}
\label{introduction}

One of the four overarching questions framing the recent report by The
Committee on the Assessment of and Outlook for Nuclear Physics is {\sl
``How does subatomic matter organize
itself?''}\,\cite{national2012Nuclear}.  This question has been at the
core of nuclear physics since Rutherford's century-old discovery of
the atomic nucleus in 1911.  The number of electrons---which equals
the number of protons in a neutral atom---determines the chemistry of
the atom. And it is this chemistry that is responsible for binding
atoms into molecules and molecules into both traditional and
fascinating new materials. But how does matter organize itself at
densities significantly higher than those found in everyday materials;
say, from $10^{4}\!-\!10^{15}\,{\rm g/cm^{3}}$.  Recall that in this
units nuclear-matter saturation density equals
$\rho_{{}_{0}}\!=\!2.48\times 10^{14}{\rm g/cm^{3}}$. Indeed, relative
to every day life these densities are so high that atoms become
pressure ionized. Understanding what novels phases of matter emerge
under these extreme conditions of density is both fascinating and
unknown. Moreover, it represents one of the grand challenges in
nuclear physics. Remarkably, most of these exotic phases---{\sl
Coulomb crystals, nuclear pasta, color superconductors}---can not be
realized under normal laboratory conditions. Yet, whereas most of
these phases have a fleeting existence in the laboratory, they attain
stability in neutron stars due to the presence of enormous
gravitational fields. In this manner neutron stars become the catalyst
for the formation of unique states of matter and provide unique
laboratories for the characterization of the ground state of cold
matter over an enormous range of densities. Note that an unavoidable
consequence of charge neutrality is that neutron-star matter is
necessarily neutron rich. This is a natural consequence of the very
low electron mass which in turn results in a high electron chemical
potential.

\begin{figure}[h]
\begin{center}
 \includegraphics[height=3.5in]{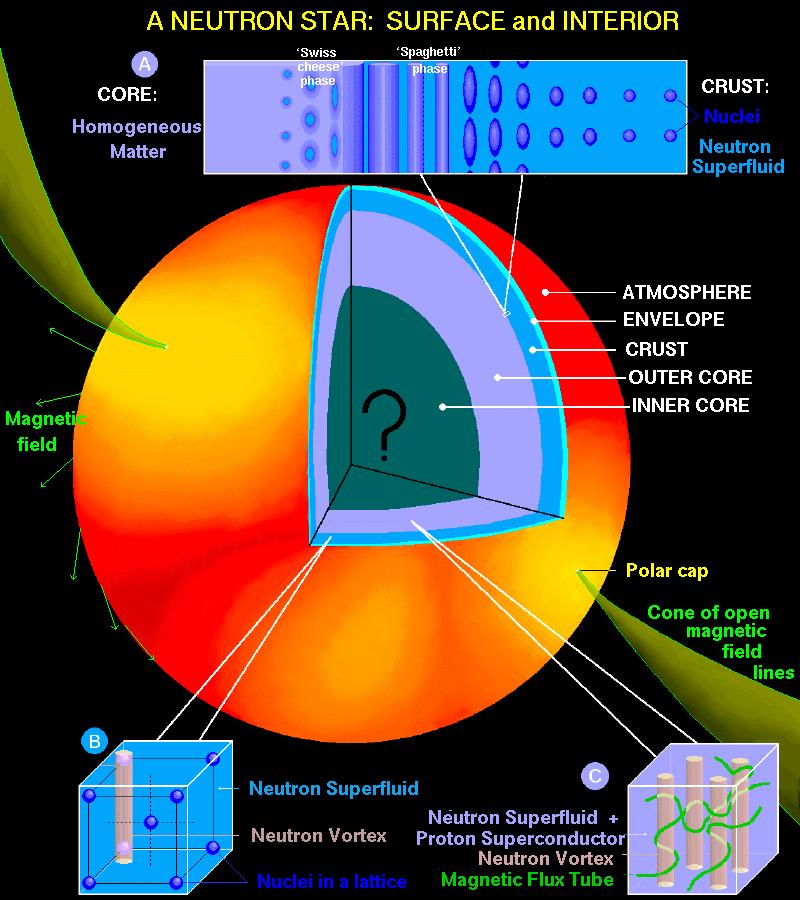}
 \vspace{-0.2cm}
 \caption{A scientifically-accurate rendition of the structure 
               and the various phases predicted to exist in a neutron 
               star (courtesy of Dany Page).}
 \label{Fig1}
\end{center}
\end{figure}

\section{Neutron-star structure}
\label{neutron-star structure} 

To appreciate the enormous dynamic range and richness displayed by
these fascinating objects we discuss briefly the anatomy of a neutron
star. For a fairly accurate rendition of the structure and phases of a
neutron star see Fig.\,\ref{Fig1}. Neutron stars contain a non-uniform
crust above a uniform liquid core that is comprised of a uniform
assembly of neutrons, protons, electrons, and muons in chemical
equilibrium and packed to densities that may exceed that of normal
nuclei by up to an order of magnitude. The highest density attained in
the stellar core depends critically on the equation of state of
neutron-rich matter, which at those high densities is poorly
constrained. However, for soft equations of state, namely, those with
a pressure that rises slowly with density, the highest density
attained at the core may be high enough for the emergence of new
exotic phases, such as pion or kaon
condensates\,\cite{Ellis:1995kz,Pons:2000xf}, strange quark
matter\,\cite{Weber:2004kj}, and color
superconductors\,\cite{Alford:1998mk,Alford:2007xm}. Nothing further
will be said in this contribution about such high-density phases.

\begin{figure}[h]
 \begin{center}
 \includegraphics[height=3.5in]{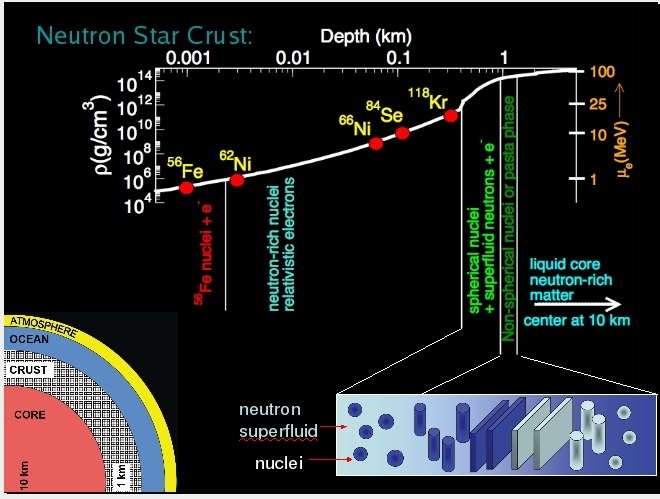}
 \vspace{-0.2cm}
 \caption{A scientifically-accurate rendition of the composition of
               the stellar crust (courtesy of Sanjay Reddy).}
 \label{Fig2}
 \end{center}
\end{figure}

\begin{figure}[ht]
\begin{center}
  \includegraphics[height=5.75cm]{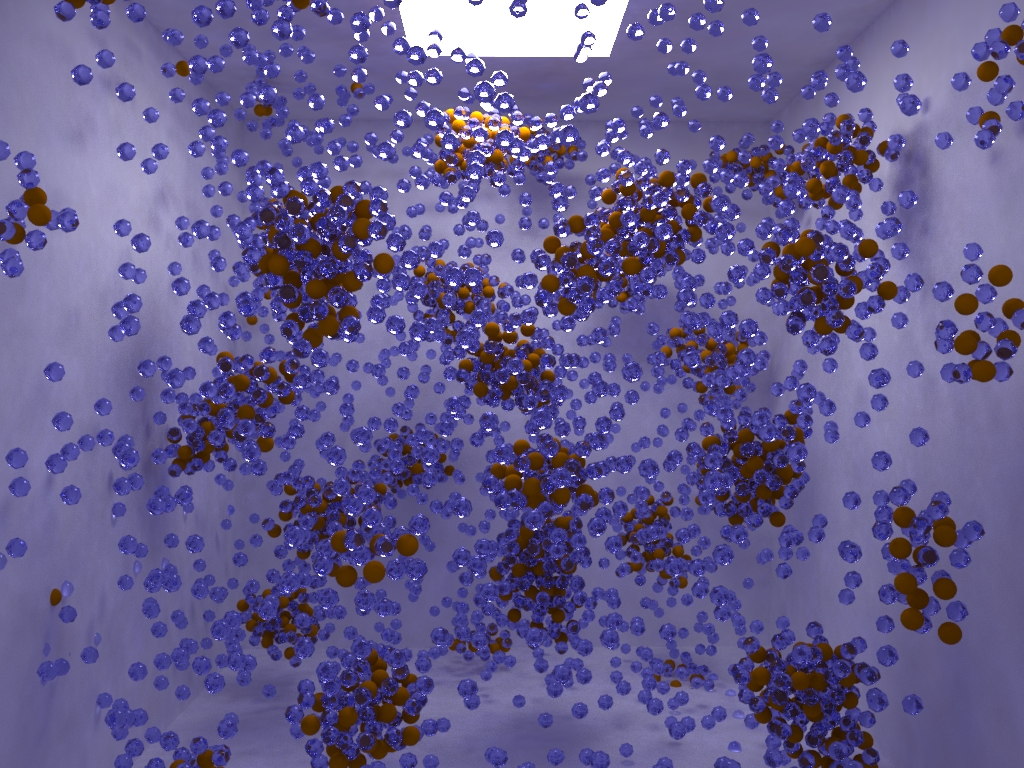} 
  \hspace{0.1cm}
  \includegraphics[height=5.75cm]{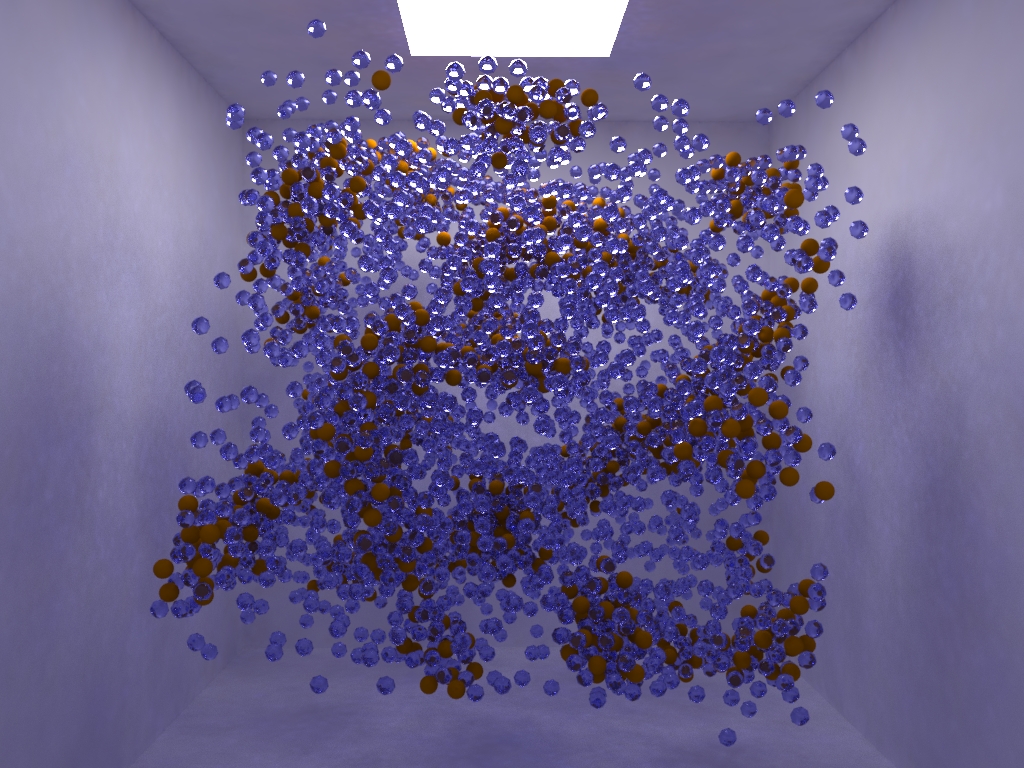} 
  \caption{(color online) Two snapshots of Monte Carlo simulations of
    neutron-rich matter, one at a density of 
    $\rho\!=\!0.01~{\rm fm}^{-3}$ (left) and the other one at 
    $\rho\!=\!0.025~{\rm fm}^{-3}$ (right), for a system of 
    4,000 nucleons at a proton fraction of
    $Y_{p}\!=\!Z/A\!=\!0.2$
    and a temperature of $T\!=\!1$~MeV.}
 \label{Fig3}
\end{center}
\end{figure}

At the other extreme, namely, at densities of about half of
nuclear-matter saturation density, the uniform core becomes unstable
against cluster formation. At these ``low'' densities the average
inter-nucleon separation increases to such an extent that it becomes
energetically favorable for the system to segregate into regions of
normal density (nuclear clusters) and regions of low density (dilute,
likely superfluid, neutron vapor).  Such a clustering instability
signals the transition from the uniform liquid core to the non-uniform
crust.  The solid crust is itself divided into an outer and an inner
region. The outer crust spans a region of about seven
orders of magnitude in density (from about $10^{4}{\rm g/cm^{3}}$ to
$4\times 10^{11}{\rm g/cm^{3}}$~\cite{Baym:1971pw,Ruester:2005fm,
RocaMaza:2008ja,RocaMaza:2011pk}).
Structurally, the outer crust is comprised of a Coulomb lattice of
neutron-rich nuclei embedded in a uniform electron gas. As the density
increases---and given that the electronic Fermi energy increases
rapidly with density---it becomes energetically favorable for
electrons to capture into protons. This results in the formation of
Coulomb crystals of progressively more neutron-rich nuclei. This
progression starts with ${}^{56}$Fe---the nucleus with the lowest
mass per nucleon---and is predicted to end with the exotic,
neutron-rich nucleus ${}^{118}$Kr (see Fig.\,\ref{Fig2}). In essence,
the most stable nucleus at a given crustal density emerges from a
competition between the electronic Fermi energy (which favors low $Z$)
and the nuclear symmetry energy (which favors $N\!\simeq\!Z$
nuclei)\,\cite{RocaMaza:2008ja,RocaMaza:2011pk}.
Eventually, however, the neutron-proton asymmetry becomes too large
for the nuclei to absorb any more neutrons and the excess neutrons go
into the formation of a dilute---likely superfluid---neutron vapor;
this signals the transition from the outer to the inner crust.  At a
{\sl neutron-drip} density of about $4\times 10^{11}{\rm g/cm^{3}}$,
${}^{118}$Kr is unable to retain any more neutrons.  As alluded
earlier, at densities approaching nuclear-matter saturation density
($\approx\!2.5\times 10^{14}{\rm g/cm^{3}}$) uniformity in the system
will be restored.  Yet the transition from the highly-ordered crystal
to the uniform liquid is both interesting and complex.  This is
because distance scales that were well separated in both the
crystalline phase (where the long-range Coulomb interaction dominates)
and in the uniform phase (where the short-range strong interaction
dominates) become comparable. This unique situation gives rise to {\sl
``Coulomb frustration''}.  Frustration, a phenomenon characterized by
the existence of a very large number of low-energy configurations,
emerges from the impossibility to simultaneously minimize all
elementary interactions in the system. Indeed, as these length scales
become comparable, competition among the elementary interactions
results in the formation of a myriad of complex structures radically
different in topology yet extremely close in energy.  Given that these
complex structures---collectively referred to as {\sl ``nuclear
pasta''}---are very close in energy, it has been speculated that the
transition from the highly ordered crystal to the uniform phase must
proceed through a series of changes in the dimensionality and topology
of these structures\,\cite{Ravenhall:1983uh,Hashimoto:1984}. Moreover,
due to the preponderance of low-energy states, frustrated systems
display an interesting and unique low-energy dynamics that has been
studied using a variety of techniques including numerical
simulations\,\cite{Horowitz:2004yf,Horowitz:2004pv,
Horowitz:2005zb,Watanabe:2003xu,Watanabe:2004tr, Watanabe:2009vi}.
In Fig.\,\ref{Fig3} we display snapshots of two such simulations at a 
density of $\rho\!=\!0.01~{\rm fm}^{-3}$---where the system still
resembles a collection of ``spherical'' clusters immersed in a dilute
neutron vapor---and at $\rho\!=\!0.025~{\rm fm}^{-3}$, where some
of the exotic shapes are starting to emerge\,\cite{Horowitz:2004yf}.

\section{Nuclear structure}
\label{nuclear structure} 

\begin{figure}[h]
\begin{center}
 \includegraphics[height=3.5in]{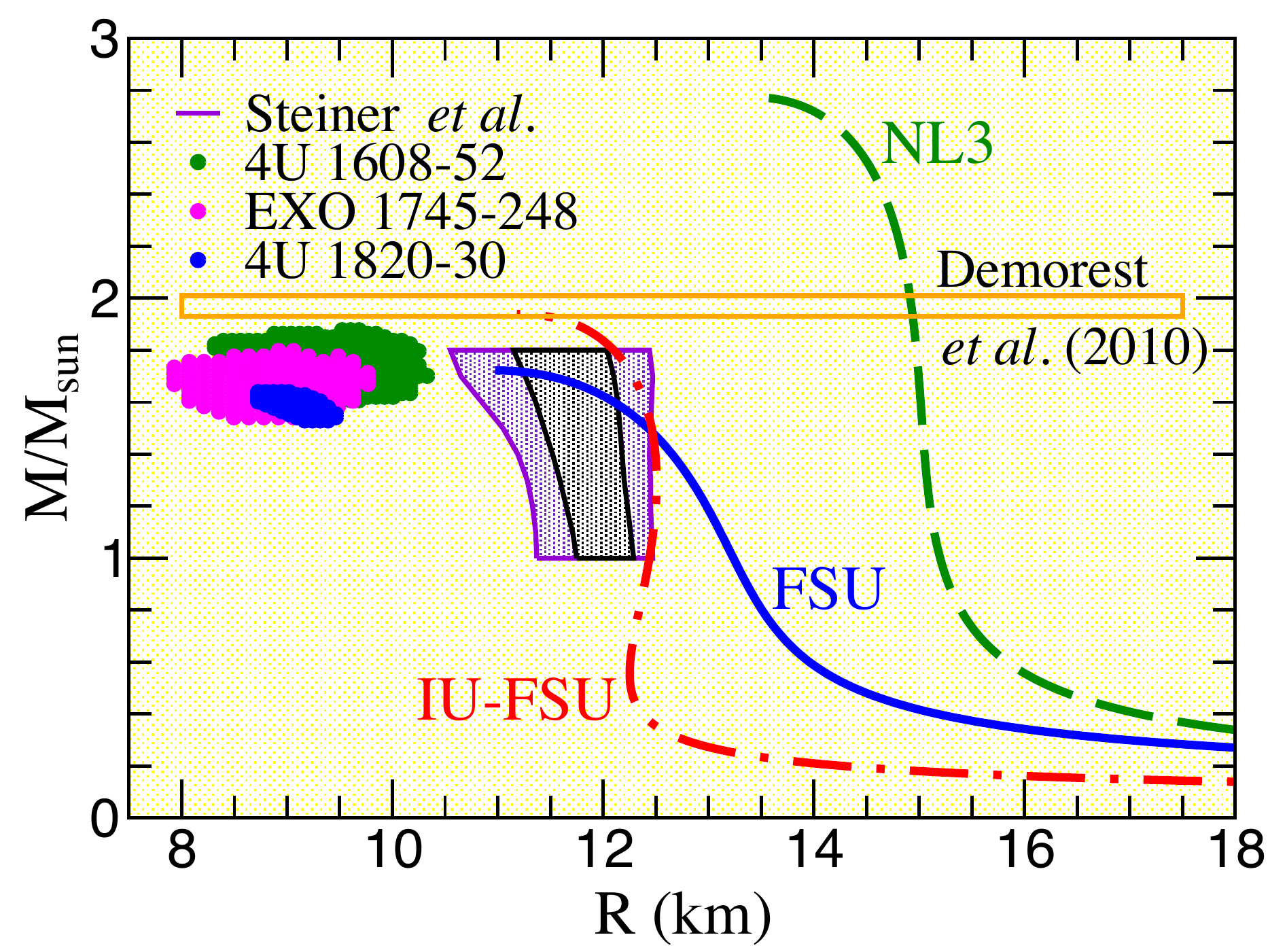}
 \vspace{-0.2cm}
 \caption{(Color online) {\sl Mass-vs-Radius} relation predicted by 
  the three relativistic mean-field models\,\cite{Fattoyev:2010mx}. The
  observational data that suggest very small stellar radii represent 
  1$\sigma$ confidence contours for the three neutron stars reported 
   in Ref.~\cite{Ozel:2010fw}. The two shaded areas that suggest larger radii
  are 1$\sigma$ and 2$\sigma$ contours extracted from the analysis of 
  Ref.~\cite{Steiner:2010fz}.}
 \label{Fig4}
\end{center}
\end{figure}

The main goal of this contribution is to invoke nuclear-structure
observables to constrain the structure, dynamics, and composition of
neutron stars.  Nuclear structure plays a critical role because in
order to prevent the collapse of the star the enormous gravitational
fields must be balanced by the pressure support of its underlying
constituents.  To illustrate this point we note that
spherically-symmetric neutron stars in hydrostatic equilibrium satisfy
the Tolman-Oppenheimer-Volkoff (TOV) equations, which are the
extension of Newton's laws to the domain of general relativity.  The
TOV equations may be expressed as a coupled set of first-order
differential equations of the following form:
 \begin{eqnarray}
   && \frac{dP}{dr}=-G\,\frac{{\cal E}(r)M(r)}{r^{2}}
         \left[1+\frac{P(r)}{{\cal E}(r)}\right]
         \left[1+\frac{4\pi r^{3}P(r)}{M(r)}\right]
         \left[1-\frac{2GM(r)}{r}\right]^{-1} \;,
         \label{TOVa}\\
   && \frac{dM}{dr}=4\pi r^{2}{\cal E}(r)\;,
         \label{TOVb}
 \label{TOV}
\end{eqnarray}
where $G$ is Newton's gravitational constant and $P(r)$, ${\cal
E}(r)$, and $M(r)$ represent the pressure, energy density, and
enclosed-mass profiles of the star, respectively. Note that the three
terms enclosed in square brackets in Eq.~(\ref{TOVa}) are of
general-relativistic origin. Notably, the only input that
neutron stars are sensitive to is the equation of state (EOS), namely,
the relation between the pressure $P$ and energy density ${\cal E}$. 
Indeed, no solution of the TOV equations is possible without a
model for the equation of state. Conversely and remarkably, each EOS
generate a unique mass-{\sl vs}-radius
relation\,\cite{Lindblom:1992}. In Fig.\,\ref{Fig4} we display 
mass-{\sl vs}-radius relations as predicted by three relativistic
mean-field models\,\cite{Fattoyev:2010mx}.  To a large extent, all
three models---NL3\,\cite{Lalazissis:1996rd, Lalazissis:1999},
FSU\,\cite{Todd-Rutel:2005fa}, and
IU-FSU\,\cite{Fattoyev:2010mx}---are able to accurately reproduce a
variety of ground-state observables (such as masses and charge radii)
throughout the nuclear chart. Yet, the predictions displayed in
Fig.\,\ref{Fig4} are significantly different. In what follows we
identify the reason for such a large model dependence and elucidate
how the measurement of certain critical laboratory observables may be
used to constrain the structure, dynamics, and composition of neutron
stars.

The starting point for the calculation of both nuclear and
neutron-star structure is the interacting Lagrangian density of
Ref.~\cite{Mueller:1996pm} supplemented by an isoscalar-isovector term
first introduced in Ref.~\cite{Horowitz:2000xj}. That is,
\begin{eqnarray}
{\mathscr L}_{\rm int} &=&
\bar\psi \left[g_{\rm s}\phi   \!-\! 
         \left(g_{\rm v}V_\mu  \!+\!
    \frac{g_{\rho}}{2}{\mbox{\boldmath $\tau$}}\cdot{\bf b}_{\mu} 
                               \!+\!    
    \frac{e}{2}(1\!+\!\tau_{3})A_{\mu}\right)\gamma^{\mu}
         \right]\psi \nonumber \\
                   &-& 
    \frac{\kappa}{3!} (g_{\rm s}\phi)^3 \!-\!
    \frac{\lambda}{4!}(g_{\rm s}\phi)^4 \!+\!
    \frac{\zeta}{4!}   g_{\rm v}^4(V_{\mu}V^\mu)^2 +
   \Lambda_{\rm v}\Big(g_{\rho}^{2}\,{\bf b}_{\mu}\cdot{\bf b}^{\mu}\Big)
                           \Big(g_{\rm v}^{2}V_{\nu}V^{\nu}\Big)\;.
 \label{LDensity}
\end{eqnarray}
The Lagrangian density includes an isodoublet nucleon field ($\psi$)
interacting via the exchange of two isoscalar mesons, a scalar
($\phi$) and a vector ($V^{\mu}$), one isovector meson ($b^{\mu}$),
and the photon ($A^{\mu}$)~\cite{Serot:1984ey,Serot:1997xg}. In
addition to meson-nucleon interactions the Lagrangian density is
supplemented by four nonlinear meson interactions with coupling
constants ($\kappa$, $\lambda$, $\zeta$, and $\Lambda_{\rm v}$)  
that are included primarily to soften the equation of state of both
symmetric nuclear matter and pure neutron matter. For a detailed
discussion on the impact of these terms on various quantities of
theoretical, experimental, and observational interest see
Ref.\,\cite{Piekarewicz:2007dx}.

Of significant relevance to the various trends displayed in
Fig.\,\ref{Fig4} are the isoscalar-vector self-interactions (scaled by
the parameter $\zeta$) and the mixed isoscalar-isovector interaction
(scaled by the parameter $\Lambda_{\rm v}$). In particular,
isoscalar-vector self-interactions may be tuned to primarily and
almost exclusively modify the equation of state of symmetric nuclear
matter at high densities. For example, M\"uller and Serot found
possible to build models with different values of $\zeta$ that
reproduce the same observed properties at saturation density, yet
predict maximum neutron star masses that may differ by almost one
solar mass\,\cite{Mueller:1996pm}. Indeed, by a fine tuning of $\zeta$
one was able to increase the maximum neutron star mass from
$1.72\,{\rm M}_{\odot}$ (in the FSU model) to $1.94\,{\rm M}_{\odot}$
(in the IU-FSU model) without adversely affecting well-known
properties of finite nuclei\,\cite{Fattoyev:2010mx}. This last value
is consistent with the recent Demorest {\sl et al.,} observation of a
$(1.97\pm0.04)\,{\rm M}_{\odot}$ neutron
star\,\cite{Demorest:2010bx}. Thus, we reach the inescapable conclusion
that the only reliable constrain on the high-density EOS of cold
nuclear matter must come from the observation of massive neutron stars.

In contrast, laboratory experiments may play a critical role in
constraining the size of neutron stars.  This is because neutron-star
radii are controlled by the density dependence of the symmetry energy
in the immediate vicinity of nuclear-matter saturation
density\,\cite{Lattimer:2006xb}.  Recall that the symmetry energy
represents the energy cost in converting protons into neutrons (or
viceversa) and may be viewed as the difference in the energy between
pure neutron matter and symmetric nuclear matter.  A particularly
critical property of the symmetry energy is its slope at saturation
density---a quantity customarily denoted by
$L$\,\cite{Piekarewicz:2008nh}. Unlike symmetric nuclear matter, the
slope of the symmetry does not vanish at saturation density.  Indeed,
$L$ is simple related to the pressure of pure neutron matter at
saturation density. That is,
\begin{equation}
  P_{0} = \frac{1}{3}\rho_{{}_{0}}L\;.
 \label{PZero}
\end {equation} 
Although the slope of the symmetry energy is not directly observable,
it is strongly correlated to the thickness of the neutron skin of
heavy nuclei\,\cite{Brown:2000,Furnstahl:2001un}.  Heavy nuclei
develop a neutron skin as a consequence of a large neutron excess 
and a Coulomb barrier that hinders the proton density at the surface of
the nucleus. The thickness of the neutron skin depends sensitively on
the pressure of neutron-rich matter: {\sl the greater the pressure the
thicker the neutron skin}. And it is exactly this same pressure that
supports neutron stars against gravitational collapse.  Thus models
with thicker neutron skins often produce neutron stars with larger
radii\,\cite{Horowitz:2000xj,Horowitz:2001ya}. Thus, it is possible to
study ``data-to-data'' relations between the neutron-rich skin of a
heavy nucleus and the radius of a neutron star.
\begin{figure}[h]
\begin{center}
 \includegraphics[width=3.5in,height=2.75in]{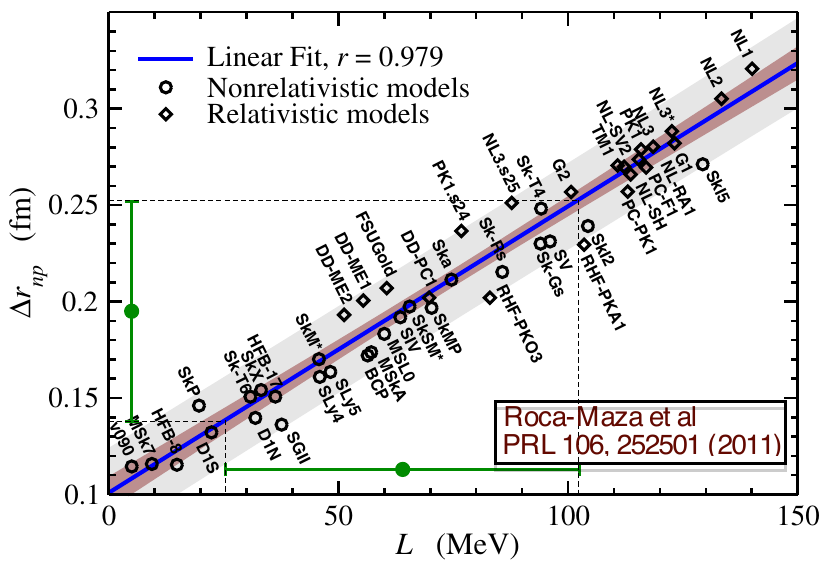}
  \hspace{0.1cm}
 \includegraphics[width=2.5in,height=2.75in]{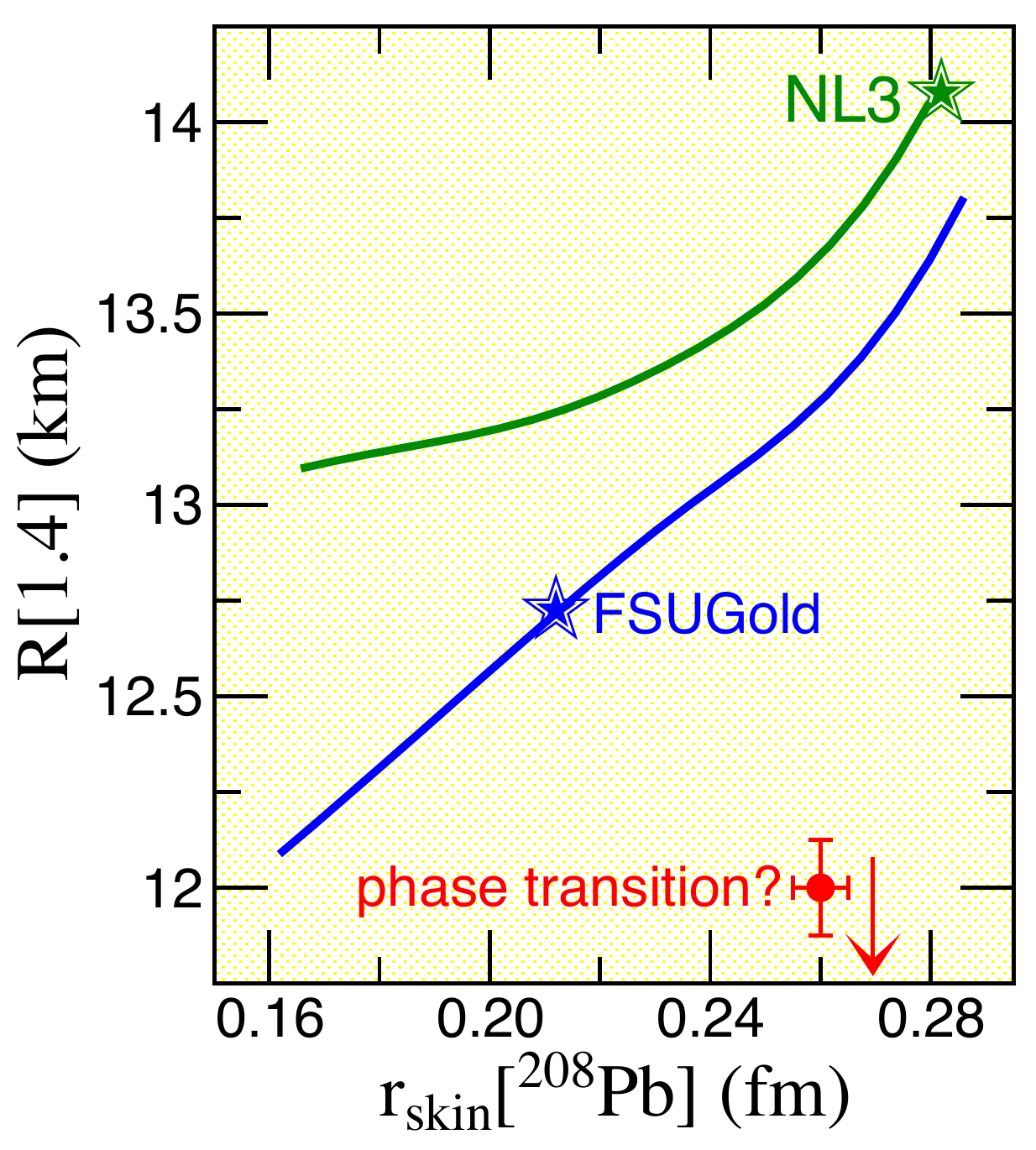}
 \vspace{-0.2cm}
 \caption{(Color online) The left-hand panel displays the correlation 
   between the neutron-skin of ${}^{208}$Pb and the slope of the 
   symmetry energy for a variety of nonrelativistic and relativistic 
   models\,\cite{RocaMaza:2011pm}. The right-hand panel shows the 
   correlation between the neutron-skin of ${}^{208}$Pb and the radius 
   of a $1.4 \,{\rm M}_{\odot}$ neutron star for two relativistic
   mean-field models.}      
 \label{Fig5}
\end{center}
\end{figure}
We illustrate these ideas in Fig.\,\ref{Fig5} where the neutron-skin
thickness of ${}^{208}$Pb ($\Delta r_{np}$) is plotted on the
left-hand panel against the slope of the symmetry energy ($L$) for a
variety of nonrelativistic and relativistic
models\,\cite{RocaMaza:2011pm}. The correlation between these two
quantities is extremely strong (0.979) and indicates that the
neutron skin of ${}^{208}$Pb may be used as a proxy for the
determination of a fundamental property of the EOS. Also shown on the
right-hand panel of Fig.\,\ref{Fig5} is a ``data-to-data'' relation
between the neutron-skin thickness of ${}^{208}$Pb and the radius of a
canonical $1.4 \,{\rm M}_{\odot}$ neutron star as predicted by
NL3\,\cite{Lalazissis:1996rd,Lalazissis:1999} and
FSU\,\cite{Todd-Rutel:2005fa}. Although the {\sl ``stars''} in the
figure indicate the predictions of these two accurately-calibrated
models, a systematic variation of the isoscalar-isovector parameter
$\Lambda_{\rm v}$ has been done to generate the two solid lines. Such
a systematic variation enables one to modify the density-dependence of
the symmetry energy without affecting well-known nuclear properties,
such as masses and charge radii. Our results establish a strong
correlation between two quantities---the neutron skin and the stellar
radius---that differ by 18 orders of magnitude! And although the
correlation is strong, it is not model independent as the radius of a
neutron star is sensitive to densities that are slightly higher than
those relevant to finite nuclei. Finally, the point labeled as {\sl
``phase transition''} is meant to indicate that a large neutron skin
in ${}^{208}$Pb accompanied by a small neutron-star radius is likely
to indicate a softening of the EOS at high densities, which may be
suggestive of a phase transition to an exotic state of matter.  Note
that although we have focus exclusively on the correlation between 
$L$ and the stellar radius, the impact of $L$ extends to a myriad 
of other neutron-star
observables\cite{Horowitz:2000xj,Horowitz:2001ya,Horowitz:2002mb,
Carriere:2002bx,Steiner:2004fi,Li:2005sr}.

\section{PREX: The Lead Radius Experiment}
\label{PREX} 

Given the instrumental role that the neutron-skin thickness of
${}^{208}$Pb plays in constraining the equation of state, the {\sl
Lead Radius EXperiment} (``PREX'') at the Jefferson Laboratory
represents a true experimental milestone. The successfully
commissioned Lead Radius Experiment has provided the first
model-independent evidence of the existence of a significant neutron
skin in ${}^{208}$Pb\,\cite{Abrahamyan:2012gp,Horowitz:2012tj}.
Building on the strength of the enormously successful parity-violating
program at the Jefferson Laboratory, PREX used parity-violating
electron scattering to provide a largely model-independent
determination of the neutron radius of $^{208}$Pb.  Parity violation
at low momentum transfers is particularly sensitive to the neutron
distribution because the neutral weak-vector boson ($Z^0$) couples
preferentially to the neutrons in the target~\cite{Donnelly:1989qs};
the coupling to the proton is suppressed by the weak mixing angle
($1\!-\!4\sin^{2}\theta_{W}\!\approx\!0.08$). Although very small,
this purely electroweak measurement may be interpreted with as much
confidence as conventional electromagnetic scattering experiments that
have been used for decades to accurately map the proton distribution.

The Lead Radius Experiment collected enough high-quality data to
provide a first constrain on the neutron radius of ${}^{208}$Pb.
Although PREX achieved the systematic control required to perform this
challenging experiment, unforeseen technical problems resulted in time
losses that significantly compromised the statistical accuracy of the
measurement. Thus, rather than achieving the original goal of a 3\%
uncertainty in the asymmetry---and a corresponding 1\% error in the
neutron radius---PREX had to settle for an error almost three times as
large. This resulted in the following value for the neutron-skin
thickness of ${}^{208}$Pb\,\cite{Abrahamyan:2012gp,Horowitz:2012tj}:
\begin{equation}
 R_{n}\!-\!R_{p} = 0.33^{+0.16}_{-0.18}~{\rm fm}.
 \label{PREX}
\end{equation}

Given that the determination of the neutron radius of a heavy nucleus 
is a problem of fundamental importance with far reaching implications 
in areas as diverse as nuclear
structure~\cite{Brown:2000,Furnstahl:2001un,Danielewicz:2003dd,
Centelles:2008vu,Centelles:2010qh}, atomic parity
violation~\cite{Pollock:1992mv,Sil:2005tg}, heavy-ion
collisions~\cite{Tsang:2004zz,Chen:2004si,Steiner:2005rd,
Shetty:2007zg,Tsang:2008fd}, and neutron-star
structure~\cite{Horowitz:2000xj,Horowitz:2001ya,Horowitz:2002mb,
Carriere:2002bx,Steiner:2004fi,Li:2005sr,Fattoyev:2010tb}, the PREX
collaboration has made a successful proposal for additional beam time
so that the original 1\% goal (or $\!\pm0.05$\,fm) may be 
attained\,\cite{PREXII:2012}.  Unfortunately, the 12-GeV upgrade of
the facility has pushed the timetable for the experiment all the way
to 2014-15. And while the scientific case for such a pivotal
experiment remains strong, the search for additional physical
observables that may be both readily accessible and strongly
correlated to the neutron skin (and thus also to $L$) is a worthwhile
enterprise. It is precisely the exploration of such a correlation
between the {\sl electric dipole polarizability} and the neutron-skin
thickness of ${}^{208}$Pb that is at the center of the next section.

\section{Pygmies and Giant Resonances}
\label{PygmiesGiants} 

A promising complementary approach to the parity-violating program
relies on the electromagnetic excitation of the electric dipole
mode~\cite{Harakeh:2001}.  For stable (medium to heavy) nuclei with a
moderate neutron excess the dipole response is concentrated on a
single fragment---{\sl the giant dipole resonance (GDR)}---that
exhausts almost 100\% of the classical Thomas-Reiche-Kunz (TRK) sum
rule.  For this mode of excitation---perceived as a collective oscillation of
neutrons against protons---the symmetry energy acts as the restoring
force.  Models with a soft symmetry energy predict large values for
the symmetry energy at the densities of relevance to the excitation of
this mode. As a consequence, the stronger restoring force of the
softer models generates a dipole response that is both hardened ({\sl
i.e.,} pushed to higher excitation energies) and quenched relative to
its stiffer counterparts.  In the particular case of the first moment
of the energy distribution, the quenching and hardening largely cancel
each other, leading to an energy-weighted sum that is---as it
should---fairly model independent. In contrast, the {\sl inverse}
energy-weighted sum, which is directly proportional to the dipole
polarizability $\alpha_{\raisebox{-1pt}{\tiny D}}$, is highly
sensitive to the density dependence of the symmetry energy, as here
the quenching and hardening act coherently~\cite{Piekarewicz:2010fa}.
Given that the neutron radius of a heavy nucleus is also sensitive to
the density dependence of the symmetry energy, the electric dipole
polarizability may be used to constrain the neutron skin. Indeed, this
sensitivity suggests the existence of the following interesting
correlation: {\sl the larger the neutron-skin thickness of
${}^{208}{\rm Pb}$, the larger its electric dipole polarizability}.

\begin{figure}[h]
\begin{center}
 \includegraphics[height=3.25in]{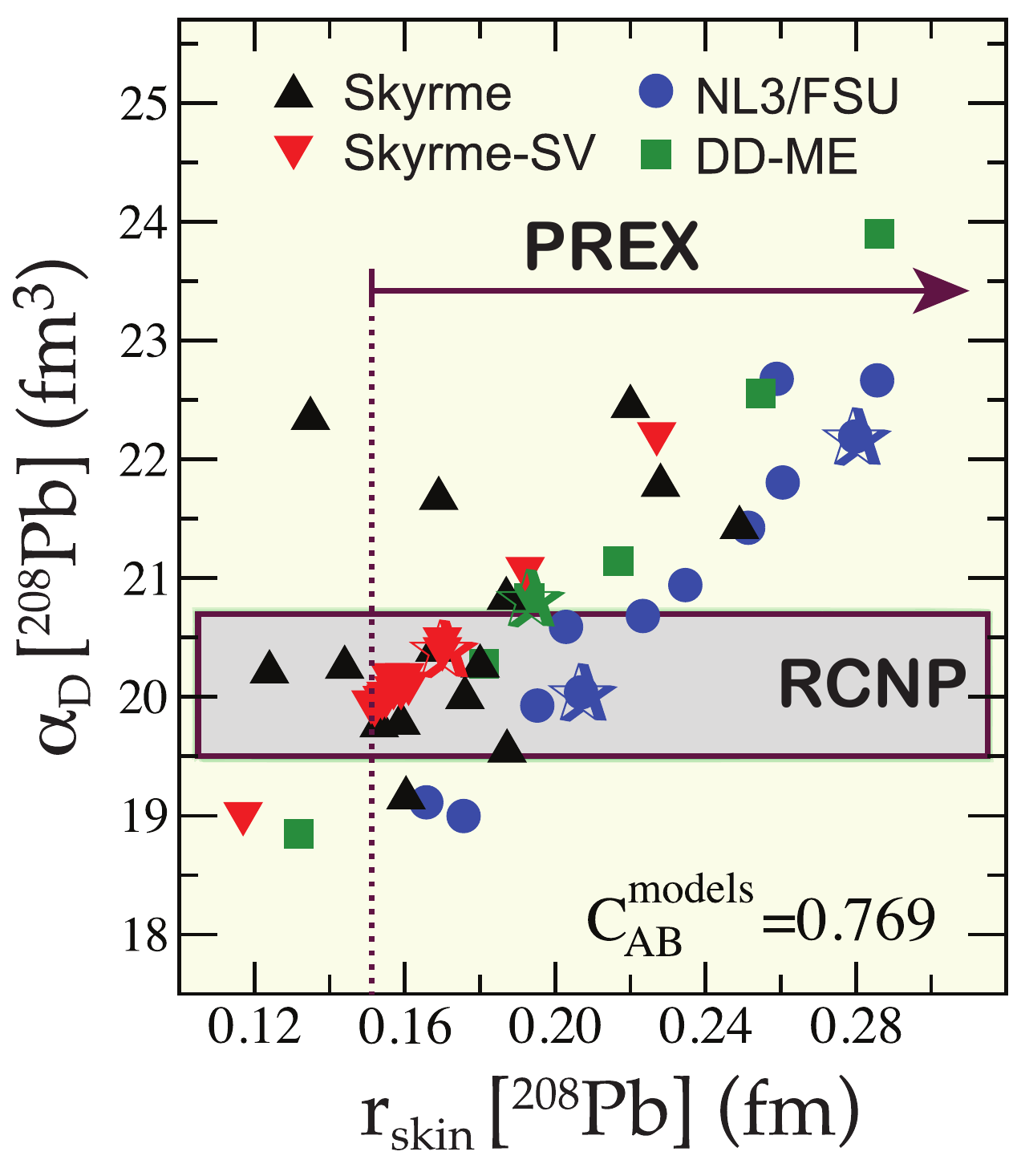}
  \hspace{0.2cm}
 \includegraphics[height=3.25in]{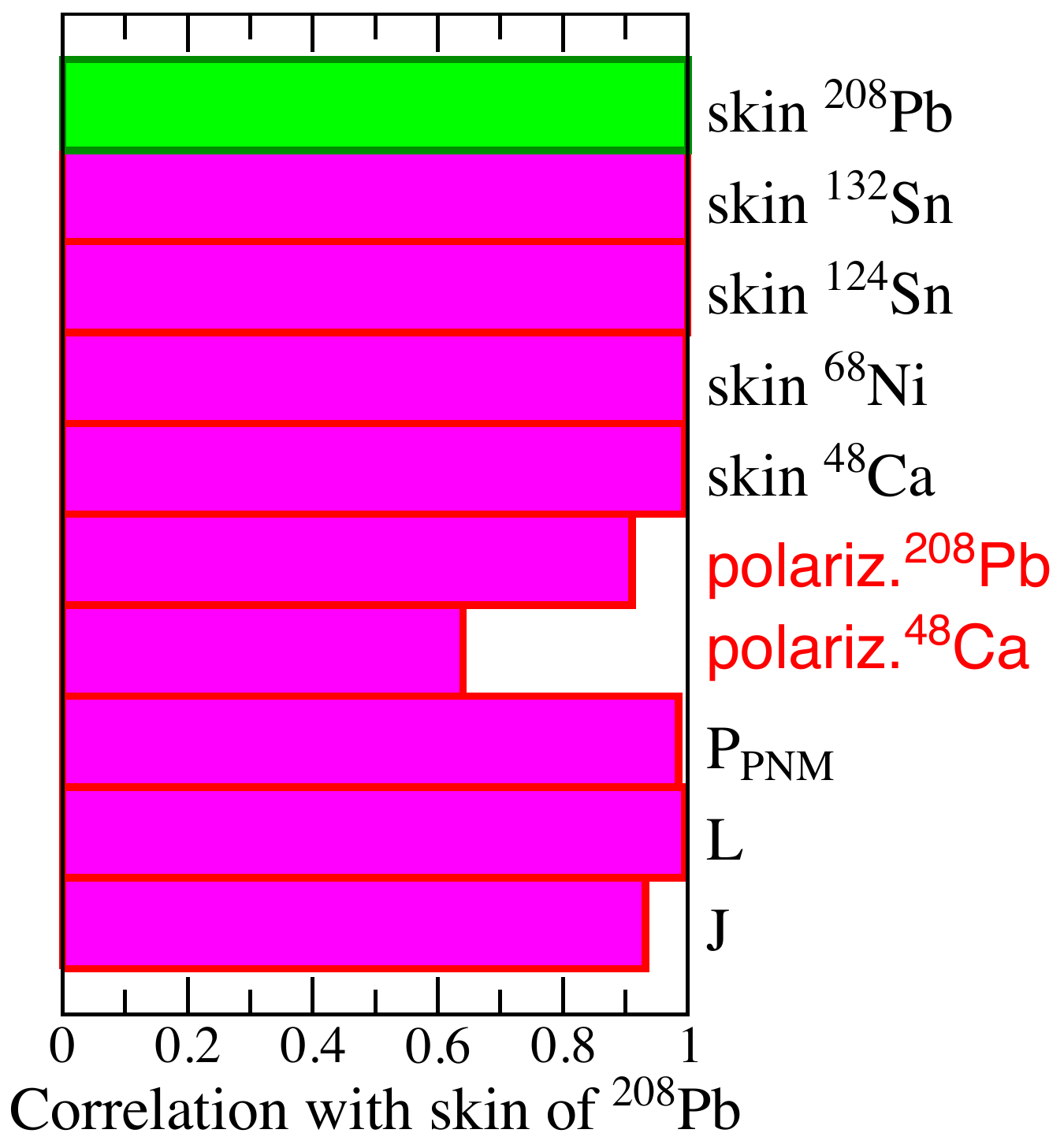}
 \vspace{-0.2cm}
 \caption{(Color online) Predictions from a variety of nuclear models
  for the electric dipole polarizability and neutron-skin thickness 
  of ${}^{208}$Pb are shown on the left-hand side of the figure. Also
  shown are constrains on the neutron-skin thickness from 
  PREX~\cite{Abrahamyan:2012gp,Horowitz:2012tj} and on the 
  dipole polarizability from RCNP~\cite{Tamii:2011pv,
  Poltoratska:2012nf}. On the right-hand side of the figure we
  show  correlation coefficients between the neutron-skin thickness 
  of ${}^{208}$Pb and several observables as obtained from a
  covariance analysis based on the FSU 
  interaction~\cite{Fattoyev:2012rm}.}
\label{Fig6}
\end{center}
\end{figure}

To test the validity of this correlation we display on the left-hand
panel of Fig.\,\ref{Fig6} the dipole polarizability in ${}^{208}{\rm
Pb}$ as a function of its corresponding neutron-skin thickness as
predicted by a large number of nuclear-structure models that have 
been calibrated to well-known properties of finite
nuclei\,\cite{Piekarewicz:2012pp}.  Once calibrated, these models
without any further adjustment are used to compute both the neutron
skin as well as the distribution of electric dipole strength. From
such a distribution of strength ($R_{\raisebox{-1pt}{\tiny E1}}$) the
dipole polarizability is readily extracted from the inverse
energy-weighted sum. That is,
\begin{equation}
 \alpha_{\raisebox{-1pt}{\tiny D}} = \frac{8\pi}{9}e^{2}
  \int_{0}^{\infty}\!\omega^{-1} 
   R_{\raisebox{-1pt}{\tiny E1}} (\omega)\,d\omega \;.
\label{AlphaD}
\end{equation}
At first glance a clear (positive) correlation between the dipole
polarizability and the neutron skin is discerned. However, on 
closer examination one observes a significant scatter in the
results---especially in the case of the standard Skyrme forces
(denoted by the black triangles).  In particular, by including the
predictions from all the 48 models under consideration, a correlation
coefficient of 0.77 was obtained.  Also shown in the figure are
experimental constraints imposed from PREX and the recent
high-resolution measurement of $\alpha_{\raisebox{-1pt}{\tiny D}}$
in $^{208}$Pb\,\cite{Tamii:2011pv,Poltoratska:2012nf}.  By imposing
these recent experimental constraints, several of the
models---especially those with either a very soft or very stiff
symmetry energy---may already be ruled out.  Evidently the correlation
between $\alpha_{\raisebox{-1pt}{\tiny D}}$ and $R_{n}\!-\!R_{p}$ is
model dependent and deserves to be investigated further.

However, to establish how the dipole polarizability may provide a
unique constraint on the neutron-skin thickness of neutron-rich nuclei
and other isovector observables we display on the right-hand panel of
Fig.\,\ref{Fig6} correlation coefficients computed using a single
underlying model, namely, FSU\,\cite{Todd-Rutel:2005fa}. For details
on the implementation of the required covariance analysis we refer the
reader to
Refs.\,\cite{Reinhard:2010wz,Fattoyev:2011ns,Fattoyev:2012rm}. According
to the model, an accurate measurement of the neutron skin-thickness in
${}^{208}{\rm Pb}$ significantly constrains the neutron skin on a
variety of other neutron-rich nuclei. Moreover, the correlation
coefficient between the neutron skin and
$\alpha_{\raisebox{-1pt}{\tiny D}}$ in ${}^{208}{\rm Pb}$ is very
large (of about 0.9). This suggests that a multi-prong approach
consisting of combined measurements of both neutron skins and
$\alpha_{\raisebox{-1pt}{\tiny D}}$---ideally on a variety of
nuclei---should significantly constrain the isovector sector of the
nuclear energy density functional as well as the EOS of neutron-rich
matter.

\begin{figure}[h]
\begin{center}
 \includegraphics[width=3.0in]{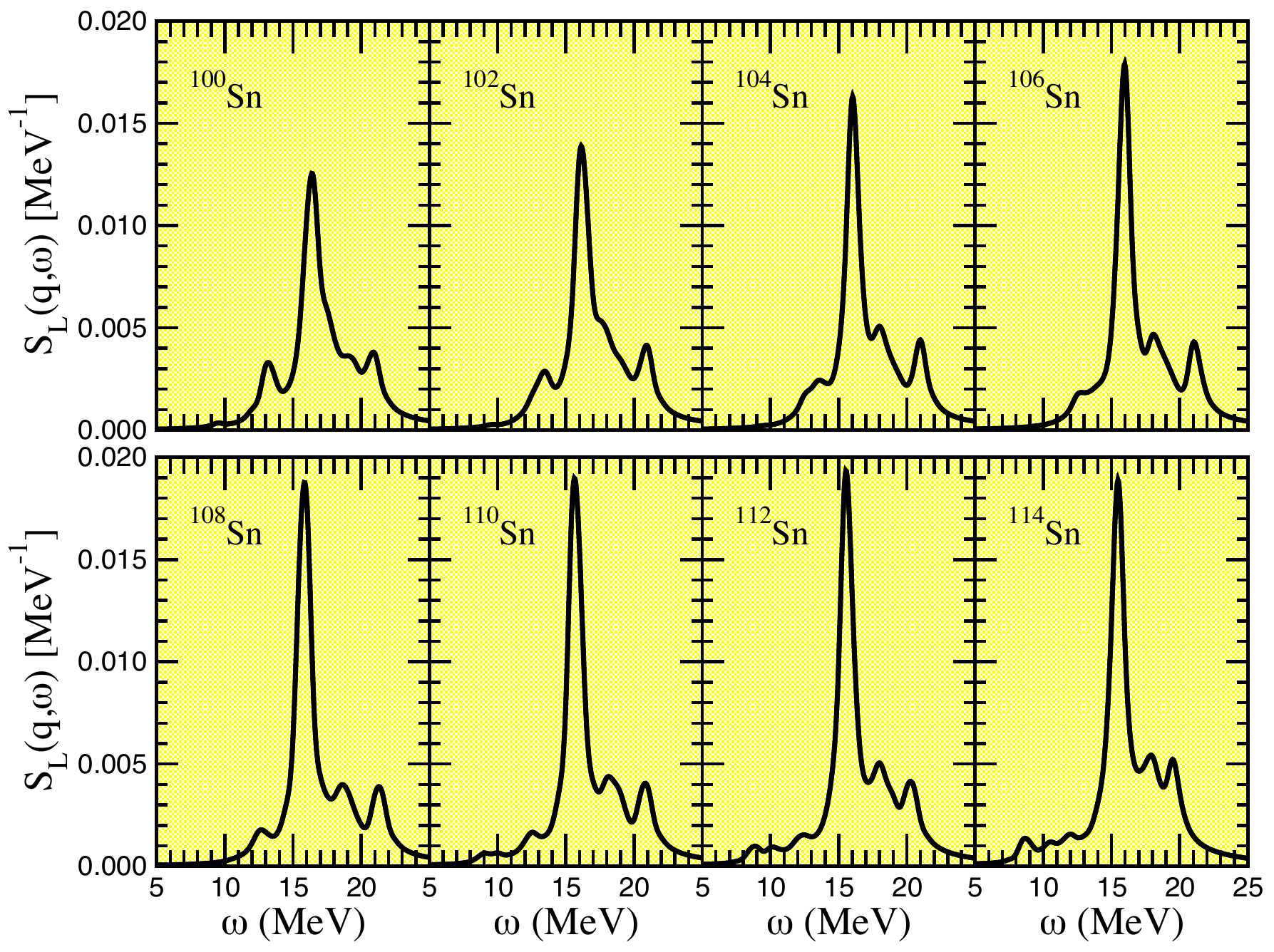}
  \hspace{0.1cm}
 \includegraphics[width=3.00in]{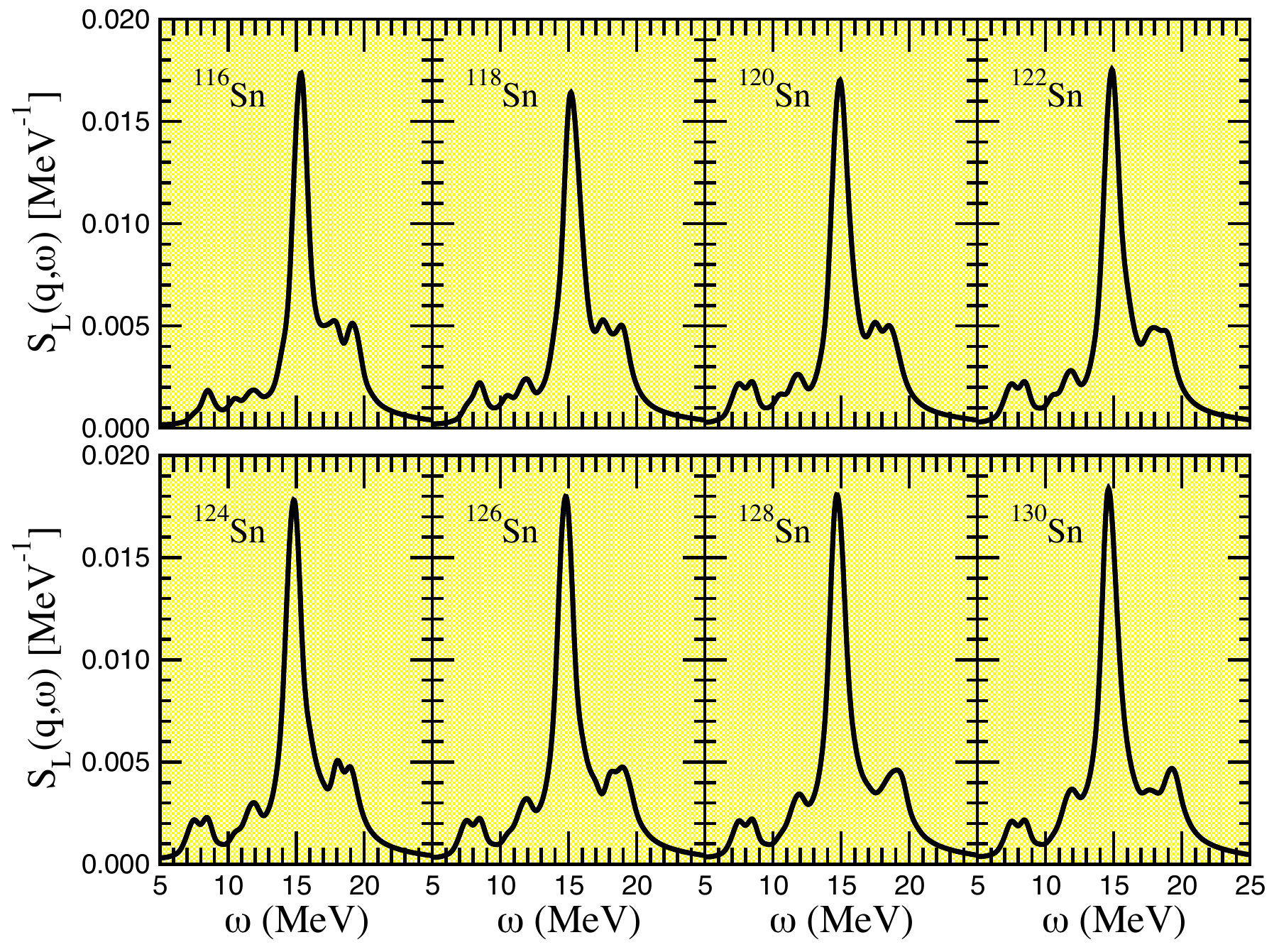}
 \vspace{-0.2cm}
 \caption{(Color online) Distribution of isovector dipole strength for all 
         neutron-even tin isotopes from ${}^{100}$Sn to ${}^{130}$Sn using
         the FSUGold parameter set\,\cite{Todd-Rutel:2005fa}. A
         detailed description of the RPA formalism required to
         generate this plot may be found in Ref.\,\cite{Piekarewicz:2006ip}.}      
 \label{Fig7}
\end{center}
\end{figure}

Naturally, a more stringent constrain on the isovector sector of the
nuclear density functional is expected to emerge along an isotopic
chain as the nucleus develops a neutron-rich skin. Concomitant with 
the development of a neutron skin one expects the emergence of low 
energy dipole strength---the so-called {\sl pygmy dipole
resonance}\,\cite{Suzuki:1990,VanIsacker:1992,Hamamoto:1996,
Hamamoto:1998,Vretenar:2000yy,Vretenar:2001hs,Paar:2004gr}.  
Thus, it has been suggested that the pygmy dipole resonance
(PDR)---speculated to be an excitation of the neutron-rich skin
against the isospin symmetric core---may be used as a constraint 
on the neutron skin of heavy nuclei\,\cite{Piekarewicz:2006ip}. In
particular, the {\sl fraction} of the energy weighted sum rule (EWSR)
exhausted by the pygmy resonance has been shown to be sensitive to the
neutron-skin thickness of heavy nuclei\,\cite{Piekarewicz:2006ip,
Tsoneva:2003gv,Tsoneva:2007fk,Klimkiewicz:2007zz,Carbone:2010az}.
Recent pioneering experiments on unstable neutron-rich isotopes in Sn, 
Sb, and Ni seem to support this assertion\,\cite{Klimkiewicz:2007zz,
Adrich:2005,Wieland:2009}. 

To illustrate these ideas we display in Fig.\,\ref{Fig7} the
distribution of isovector dipole strength for all even-even
Sn-isotopes from ${}^{100}$Sn up to ${}^{130}$Sn.  The large
collective structure in the $\omega\!\sim\!15\!-\!16$~MeV region
represents the isovector giant dipole resonance. For medium-to-heavy
nuclei this collective vibration represents a coherent oscillation of
all protons against all neutrons and is well-developed along the whole
isotopic chain~\cite{Harakeh:2001,Bertsch:1994}.  As is characteristic
of these collective excitations, a large fraction of the
energy-weighted sum rule is exhausted by this one resonance. But
certainly not all! The development of low-energy
($\omega\!\sim\!7\!-\!9$~MeV) dipole strength with increasing neutron
number is clearly discerned. Indeed, the progressive addition of
neutrons results in both the emergence of a neutron-rich skin and a
well developed, albeit small, low-energy resonance.

\begin{figure}[h]
\begin{center}
 \includegraphics[height=3.25in]{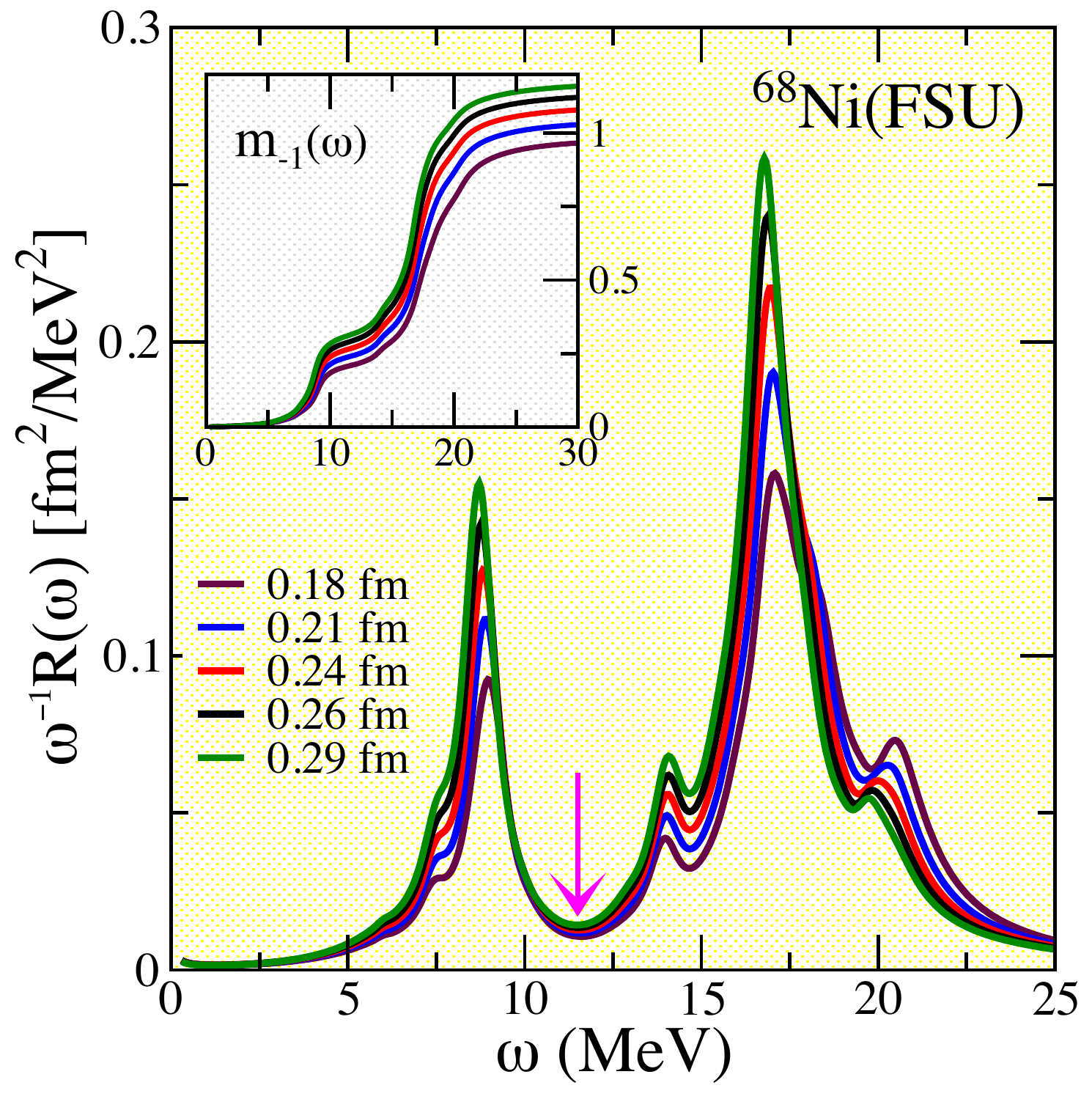}
  \hspace{0.1cm}
 \includegraphics[height=3.25in]{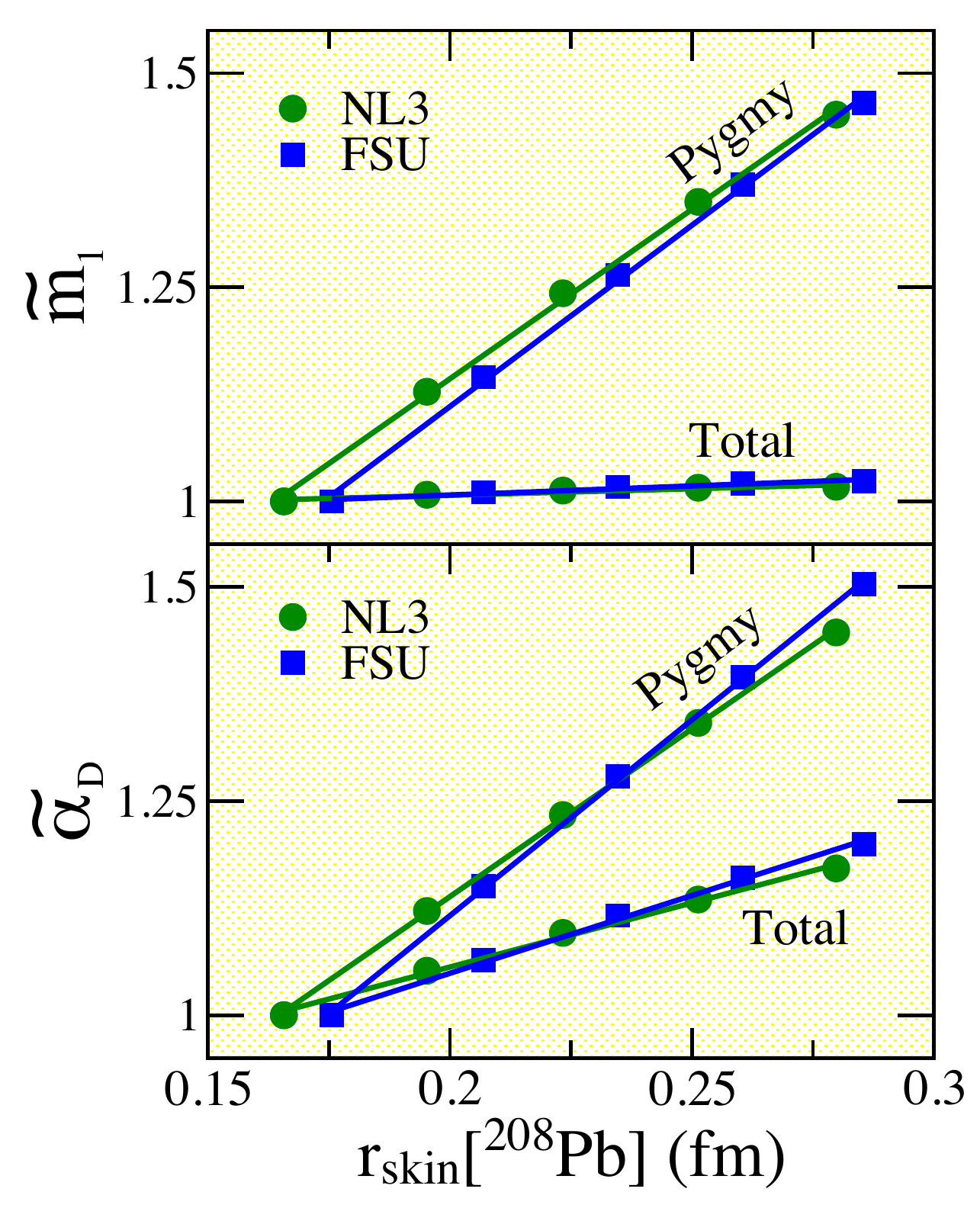}
 \vspace{-0.2cm}
 \caption{(Color online) The inverse energy weighted dipole response
   in ${}^{68}$Ni computed with the FSU family of effective interactions
   is shown on the left-hand side of the figure. The inset displays the 
   cumulative sum as defined in Eq.\,(\ref{mm1Def}). The arrow
   indicates the (ad-hoc) energy at which the low-energy (pygmy)
   response is separated from the high-energy (giant) response. 
   On the right-hand side the fractional change in the energy weighted 
   sum and dipole polarizability for ${}^{68}$Ni are displayed as a
   function of the neutron-skin thickness of ${}^{208}$Pb. See 
   Ref.\,\cite{Piekarewicz:2010fa} for more details.}
\label{Fig8}
\end{center}
\end{figure}

But if the fraction of the {\sl energy weighted sum rule} exhausted by
the pygmy resonance has been shown to be sensitive to the neutron skin
of heavy nuclei, the fraction of the {\sl inverse energy weighted sum
rule} carried by the PDR appears to be even more sensitive. Again,
this is related to the fact that the hardening and quenching of the
isovector dipole response is more extreme for models with a soft
symmetry energy. The inverse energy weighted response $\omega^{-1}
R(\omega)$ is displayed on the left-hand panel of
Fig.\,\ref{Fig8}. Given that the $\omega^{-1}$ factor enhances
preferentially the low-energy part of response, the Pygmy resonance
accounts for a significant fraction (of about 20-25\%) of the $m_{-1}$
moment, which is directly related to the dipole polarizability through
Eq.\,(\ref{AlphaD}). This should be contrasted against the EWSR where
the Pygmy resonance exhausts merely 5-8\% of the total
sum\,\cite{Piekarewicz:2010fa}. Moreover, the inverse energy weighting
enhances further the response generated from models with a stiff
symmetry energy. Pictorially, this behavior is best illustrated in the
inset of Fig.\,\ref{Fig8} which displays the {\sl cumulative}
$m_{-1}(\omega)$ sum:
\begin{equation}
   m_{-1}(\omega) = \int_{0}^{\omega}\frac{R(\omega')}{\omega'} d\omega'\;.
 \label{mm1Def}
\end{equation}
The inset provides a clear indication that both the total $m_{-1}$
moment as well as the fraction contained in the Pygmy resonance are
highly sensitive to the neutron-skin thickness of ${}^{208}$Pb.  To
heighten this sensitivity we display on the right-hand panel of
Fig.\,\ref{Fig8} the {\sl fractional change} in both the total and
Pygmy contributions to the $m_{1}$ moment and to the dipole
polarizability $\alpha_{D}$ as a function of the neutron skin of
${}^{208}$Pb (we denote these fractional changes with a {\sl
``tilde''} in the figure). These results illustrate the strong
correlation between the neutron skin and $\alpha_{D}$ and establish
how a combined measurement of these laboratory observables will be of
vital importance in constraining the isovector sector of the nuclear
density functional.

\section{Conclusions}
\label{conclusions} 
 
Measurements of neutron radii provide important constraints on the
isovector sector of nuclear density functionals and offer vital
guidance in areas as diverse as atomic parity violation, heavy-ion
collisions, and neutron-star structure. In this contribution we
examined the possibility of using the quintessential nuclear
mode---the isovector dipole resonance---as a promising 
complementary observable.
For this mode of excitation in which protons oscillate coherently
against neutrons, the symmetry energy acts as its restoring force.
Thus, models with a soft symmetry energy predict large values for the
symmetry energy at the densities of relevance to the excitation of
this mode. As a consequence, softer models generates a dipole response
that is both hardened and quenched relative to the stiffer
models. However, being protected by the TRK sum rule, the energy
weighted sum rule is largely insensitive to this behavior. In
contrast, for the inverse energy-weighted sum---which is
directly proportional to the electric dipole polarizability
$\alpha_{\raisebox{-1pt}{\tiny D}}$---the quenching and hardening act
in tandem.  Thus, models with a soft symmetry energy predict smaller
values of $\alpha_{\raisebox{-1pt}{\tiny D}}$ than their stiffer
counterparts. This results in a powerful ``data-to-data'' relation:
{\sl the smaller $\alpha_{\raisebox{-1pt}{\tiny D}}$ the thinner the
neutron skin}.

A particular intriguing question concerns the role of the pygmy dipole
resonance in constraining the density dependence of the symmetry
energy.  Regarded as an oscillation of the neutron-rich skin of a
heavy nucleus against its isospin-symmetric core, the PDR was
suggested to be strongly correlated to the neutron skin. In the
particular case of the Tin isotopes, a clear emergence of low-energy
dipole strength is observed as the nucleus develops a neutron-rich
skin. Moreover, it appears that although the total EWSR is fairly
insensitive to the density dependence of the symmetry energy, the
fraction of the EWSR exhausted by the pygmy displays some
sensitivity. However, in the case of the dipole polarizability the
conclusion that the PDR is highly sensitive to the density dependence
of symmetry energy appears inescapable. Indeed, in the particular case
of ${}^{68}$Ni the PDR accounts for 20-25\% of the total dipole
polarizability and displays a strong sensitivity to the neutron
skin. Yet, many open questions remain.
First and foremost, the strong correlation between the PDR and the
neutron skin found here appears to be model dependent. While we
support the notion of a strong correlation between these two
observables, Reinhard and Nazarewicz conclude that the neutron-skin
thickness of ${}^{208}$Pb is very weakly correlated to the low-energy
dipole strength\,\cite{Reinhard:2010wz}. Moreover, even the nature of
the low-energy mode is unclear. Is it indeed a collective mode? Is it a skin
oscillation? Can it be cleanly decoupled from the low-energy tail of
the giant resonance? Although most of these issues were not addressed
in this contribution, attempts to answer some of these question may be found
in two recent reviews\,\cite{Paar:2007bk,Paar:2010ww}. Regardless of
the nature of the mode, the emergence of low-energy dipole strength as
nuclei develop a neutron-rich skin is an incontrovertible fact. As
such, it should play a pivotal role in constraining the EOS of
neutron-rich matter.

In summary, motivated by two seminal
experiments\,\cite{Abrahamyan:2012gp,Tamii:2011pv}, we examined
possible correlations between the electric dipole polarizability and
the neutron skin of neutron-rich nuclei. The neutron-skin thickness of
a heavy nucleus is a quantity of critical importance for our
understanding of a variety of nuclear and astrophysical phenomena.  In
particular, the neutron-skin thickness of $^{208}$Pb can provide
stringent constrains on the density dependence of the symmetry energy
which, in turn, has a strong impact on the structure, dynamics, and
composition of neutron stars.  We conclude that precise measurements
of neutron skins and $\alpha_{\raisebox{-1pt}{\tiny D}}$---ideally on
a variety of nuclei--- should significantly constrain the isovector
sector of the nuclear energy density functional and will provide
critical insights into the nature of neutron-rich matter.

\section*{Acknowledgments}
  This work was supported in part by grant DE-FD05-92ER40750
  from the U.S. Department of Energy.


\section*{References}
\begin{thebibliography}{80}
\expandafter\ifx\csname natexlab\endcsname\relax\def\natexlab#1{#1}\fi
\expandafter\ifx\csname bibnamefont\endcsname\relax
  \def\bibnamefont#1{#1}\fi
\expandafter\ifx\csname bibfnamefont\endcsname\relax
  \def\bibfnamefont#1{#1}\fi
\expandafter\ifx\csname citenamefont\endcsname\relax
  \def\citenamefont#1{#1}\fi
\expandafter\ifx\csname url\endcsname\relax
  \def\url#1{\texttt{#1}}\fi
\expandafter\ifx\csname urlprefix\endcsname\relax\def\urlprefix{URL }\fi
\providecommand{\bibinfo}[2]{#2}
\providecommand{\eprint}[2][]{\url{#2}}

\bibitem{national2012Nuclear}
\bibinfo{author}{\bibfnamefont{The Committee on the Assessment of 
 and Outlook for Nuclear Physics; Board on Physics and Astronomy; 
 Division on Engineering and Physical Sciences; National Research Council}},
 \emph{\bibinfo{title}{Nuclear Physics: Exploring the Heart of Matter}}
  (\bibinfo{publisher}{The National Academies Press}, \bibinfo{year}{2012}),
  ISBN \bibinfo{isbn}{9780309260404},
  \urlprefix\url{http://www.nap.edu/openbook.php?record\_id=13438}.

\bibitem{Ellis:1995kz}
\bibinfo{author}{\bibfnamefont{P.~J.} \bibnamefont{Ellis}},
  \bibinfo{author}{\bibfnamefont{R.}~\bibnamefont{Knorren}}, \bibnamefont{and}
  \bibinfo{author}{\bibfnamefont{M.}~\bibnamefont{Prakash}},
  \bibinfo{journal}{Phys.Lett.} \textbf{\bibinfo{volume}{B349}},
  \bibinfo{pages}{11} (\bibinfo{year}{1995}).

\bibitem{Pons:2000xf}
\bibinfo{author}{\bibfnamefont{J.~A.} \bibnamefont{Pons}},
  \bibinfo{author}{\bibfnamefont{J.~A.} \bibnamefont{Miralles}},
  \bibinfo{author}{\bibfnamefont{M.}~\bibnamefont{Prakash}}, \bibnamefont{and}
  \bibinfo{author}{\bibfnamefont{J.~M.} \bibnamefont{Lattimer}},
  \bibinfo{journal}{Astrophys. J.} \textbf{\bibinfo{volume}{553}},
  \bibinfo{pages}{382} (\bibinfo{year}{2001}).

\bibitem{Weber:2004kj}
\bibinfo{author}{\bibfnamefont{F.}~\bibnamefont{Weber}},
  \bibinfo{journal}{Prog. Part. Nucl. Phys.} \textbf{\bibinfo{volume}{54}},
  \bibinfo{pages}{193} (\bibinfo{year}{2005}).

\bibitem{Alford:1998mk}
\bibinfo{author}{\bibfnamefont{M.~G.} \bibnamefont{Alford}},
  \bibinfo{author}{\bibfnamefont{K.}~\bibnamefont{Rajagopal}},
  \bibnamefont{and} \bibinfo{author}{\bibfnamefont{F.}~\bibnamefont{Wilczek}},
  \bibinfo{journal}{Nucl. Phys.} \textbf{\bibinfo{volume}{B537}},
  \bibinfo{pages}{443} (\bibinfo{year}{1999}).

\bibitem{Alford:2007xm}
\bibinfo{author}{\bibfnamefont{M.~G.} \bibnamefont{Alford}},
  \bibinfo{author}{\bibfnamefont{A.}~\bibnamefont{Schmitt}},
  \bibinfo{author}{\bibfnamefont{K.}~\bibnamefont{Rajagopal}},
  \bibnamefont{and} \bibinfo{author}{\bibfnamefont{T.}~\bibnamefont{Schafer}},
  \bibinfo{journal}{Rev. Mod. Phys.} \textbf{\bibinfo{volume}{80}},
  \bibinfo{pages}{1455} (\bibinfo{year}{2008}).

\bibitem{Baym:1971pw}
\bibinfo{author}{\bibfnamefont{G.}~\bibnamefont{Baym}},
  \bibinfo{author}{\bibfnamefont{C.}~\bibnamefont{Pethick}}, \bibnamefont{and}
  \bibinfo{author}{\bibfnamefont{P.}~\bibnamefont{Sutherland}},
  \bibinfo{journal}{Astrophys. J.} \textbf{\bibinfo{volume}{170}},
  \bibinfo{pages}{299} (\bibinfo{year}{1971}).

\bibitem{Ruester:2005fm}
\bibinfo{author}{\bibfnamefont{S.~B.} \bibnamefont{Ruester}},
  \bibinfo{author}{\bibfnamefont{M.}~\bibnamefont{Hempel}}, \bibnamefont{and}
  \bibinfo{author}{\bibfnamefont{J.}~\bibnamefont{Schaffner-Bielich}},
  \bibinfo{journal}{Phys. Rev.} \textbf{\bibinfo{volume}{C73}},
  \bibinfo{pages}{035804} (\bibinfo{year}{2006}).

\bibitem{RocaMaza:2008ja}
\bibinfo{author}{\bibfnamefont{X.}~\bibnamefont{Roca-Maza}} \bibnamefont{and}
  \bibinfo{author}{\bibfnamefont{J.}~\bibnamefont{Piekarewicz}},
  \bibinfo{journal}{Phys. Rev.} \textbf{\bibinfo{volume}{C78}},
  \bibinfo{pages}{025807} (\bibinfo{year}{2008}).

\bibitem{RocaMaza:2011pk}
\bibinfo{author}{\bibfnamefont{X.}~\bibnamefont{Roca-Maza}},
  \bibinfo{author}{\bibfnamefont{J.}~\bibnamefont{Piekarewicz}},
  \bibinfo{author}{\bibfnamefont{T.}~\bibnamefont{Garcia-Galvez}},
  \bibnamefont{and} \bibinfo{author}{\bibfnamefont{M.}~\bibnamefont{Centelles}}
  (\bibinfo{year}{2011}{\natexlab{a}}), \bibinfo{note}{contribution to the book
  {\sl Neutron Star Crust} (Nova Publishers)}, \eprint{1109.3011}.

\bibitem{Ravenhall:1983uh}
\bibinfo{author}{\bibfnamefont{D.~G.} \bibnamefont{Ravenhall}},
  \bibinfo{author}{\bibfnamefont{C.~J.} \bibnamefont{Pethick}},
  \bibnamefont{and} \bibinfo{author}{\bibfnamefont{J.~R.}
  \bibnamefont{Wilson}}, \bibinfo{journal}{Phys. Rev. Lett.}
  \textbf{\bibinfo{volume}{50}}, \bibinfo{pages}{2066} (\bibinfo{year}{1983}).

\bibitem{Hashimoto:1984}
\bibinfo{author}{\bibfnamefont{M.}~\bibnamefont{Hashimoto}},
  \bibinfo{author}{\bibfnamefont{H.}~\bibnamefont{Seki}}, \bibnamefont{and}
  \bibinfo{author}{\bibfnamefont{M.}~\bibnamefont{Yamada}},
  \bibinfo{journal}{Prog. Theor. Phys.} \textbf{\bibinfo{volume}{71}},
  \bibinfo{pages}{320} (\bibinfo{year}{1984}).

\bibitem{Horowitz:2004yf}
\bibinfo{author}{\bibfnamefont{C.~J.} \bibnamefont{Horowitz}},
  \bibinfo{author}{\bibfnamefont{M.~A.} \bibnamefont{Perez-Garcia}},
  \bibnamefont{and}
  \bibinfo{author}{\bibfnamefont{J.}~\bibnamefont{Piekarewicz}},
  \bibinfo{journal}{Phys. Rev.} \textbf{\bibinfo{volume}{C69}},
  \bibinfo{pages}{045804} (\bibinfo{year}{2004}{\natexlab{a}}).

\bibitem{Horowitz:2004pv}
\bibinfo{author}{\bibfnamefont{C.~J.} \bibnamefont{Horowitz}},
  \bibinfo{author}{\bibfnamefont{M.~A.} \bibnamefont{Perez-Garcia}},
  \bibinfo{author}{\bibfnamefont{J.}~\bibnamefont{Carriere}},
  \bibinfo{author}{\bibfnamefont{D.~K.} \bibnamefont{Berry}}, \bibnamefont{and}
  \bibinfo{author}{\bibfnamefont{J.}~\bibnamefont{Piekarewicz}},
  \bibinfo{journal}{Phys. Rev.} \textbf{\bibinfo{volume}{C70}},
  \bibinfo{pages}{065806} (\bibinfo{year}{2004}{\natexlab{b}}).

\bibitem{Horowitz:2005zb}
\bibinfo{author}{\bibfnamefont{C.~J.} \bibnamefont{Horowitz}},
  \bibinfo{author}{\bibfnamefont{M.~A.} \bibnamefont{Perez-Garcia}},
  \bibinfo{author}{\bibfnamefont{D.~K.} \bibnamefont{Berry}}, \bibnamefont{and}
  \bibinfo{author}{\bibfnamefont{J.}~\bibnamefont{Piekarewicz}},
  \bibinfo{journal}{Phys. Rev.} \textbf{\bibinfo{volume}{C72}},
  \bibinfo{pages}{035801} (\bibinfo{year}{2005}).

\bibitem{Watanabe:2003xu}
\bibinfo{author}{\bibfnamefont{G.}~\bibnamefont{Watanabe}},
  \bibinfo{author}{\bibfnamefont{K.}~\bibnamefont{Sato}},
  \bibinfo{author}{\bibfnamefont{K.}~\bibnamefont{Yasuoka}}, \bibnamefont{and}
  \bibinfo{author}{\bibfnamefont{T.}~\bibnamefont{Ebisuzaki}},
  \bibinfo{journal}{Phys. Rev.} \textbf{\bibinfo{volume}{C68}},
  \bibinfo{pages}{035806} (\bibinfo{year}{2003}).

\bibitem{Watanabe:2004tr}
\bibinfo{author}{\bibfnamefont{G.}~\bibnamefont{Watanabe}},
  \bibinfo{author}{\bibfnamefont{T.}~\bibnamefont{Maruyama}},
  \bibinfo{author}{\bibfnamefont{K.}~\bibnamefont{Sato}},
  \bibinfo{author}{\bibfnamefont{K.}~\bibnamefont{Yasuoka}}, \bibnamefont{and}
  \bibinfo{author}{\bibfnamefont{T.}~\bibnamefont{Ebisuzaki}},
  \bibinfo{journal}{Phys. Rev. Lett.} \textbf{\bibinfo{volume}{94}},
  \bibinfo{pages}{031101} (\bibinfo{year}{2005}).

\bibitem{Watanabe:2009vi}
\bibinfo{author}{\bibfnamefont{G.}~\bibnamefont{Watanabe}},
  \bibinfo{author}{\bibfnamefont{H.}~\bibnamefont{Sonoda}},
  \bibinfo{author}{\bibfnamefont{T.}~\bibnamefont{Maruyama}},
  \bibinfo{author}{\bibfnamefont{K.}~\bibnamefont{Sato}},
  \bibinfo{author}{\bibfnamefont{K.}~\bibnamefont{Yasuoka}},
  \bibnamefont{et~al.}, \bibinfo{journal}{Phys. Rev. Lett.}
  \textbf{\bibinfo{volume}{103}}, \bibinfo{pages}{121101}
  (\bibinfo{year}{2009}).

\bibitem{Lindblom:1992}
\bibinfo{author}{\bibfnamefont{L.}~\bibnamefont{{Lindblom}}},
  \bibinfo{journal}{Astrophys. J.} \textbf{\bibinfo{volume}{398}}, \bibinfo{pages}{569}
  (\bibinfo{year}{1992}).

\bibitem{Fattoyev:2010mx}
\bibinfo{author}{\bibfnamefont{F.~J.} \bibnamefont{Fattoyev}},
  \bibinfo{author}{\bibfnamefont{C.~J.} \bibnamefont{Horowitz}},
  \bibinfo{author}{\bibfnamefont{J.}~\bibnamefont{Piekarewicz}},
  \bibnamefont{and} \bibinfo{author}{\bibfnamefont{G.}~\bibnamefont{Shen}},
  \bibinfo{journal}{Phys. Rev.} \textbf{\bibinfo{volume}{C82}},
  \bibinfo{pages}{055803} (\bibinfo{year}{2010}).

\bibitem{Lalazissis:1996rd}
\bibinfo{author}{\bibfnamefont{G.~A.} \bibnamefont{Lalazissis}},
  \bibinfo{author}{\bibfnamefont{J.}~\bibnamefont{Konig}}, \bibnamefont{and}
  \bibinfo{author}{\bibfnamefont{P.}~\bibnamefont{Ring}},
  \bibinfo{journal}{Phys. Rev.} \textbf{\bibinfo{volume}{C55}},
  \bibinfo{pages}{540} (\bibinfo{year}{1997}).

\bibitem{Lalazissis:1999}
\bibinfo{author}{\bibfnamefont{G.~A.} \bibnamefont{Lalazissis}},
  \bibinfo{author}{\bibfnamefont{S.}~\bibnamefont{Raman}}, \bibnamefont{and}
  \bibinfo{author}{\bibfnamefont{P.}~\bibnamefont{Ring}}, \bibinfo{journal}{At.
  Data Nucl. Data Tables} \textbf{\bibinfo{volume}{71}}, \bibinfo{pages}{1}
  (\bibinfo{year}{1999}).

\bibitem{Todd-Rutel:2005fa}
\bibinfo{author}{\bibfnamefont{B.~G.} \bibnamefont{Todd-Rutel}}
  \bibnamefont{and}
  \bibinfo{author}{\bibfnamefont{J.}~\bibnamefont{Piekarewicz}},
  \bibinfo{journal}{Phys. Rev. Lett} \textbf{\bibinfo{volume}{95}},
  \bibinfo{pages}{122501} (\bibinfo{year}{2005}).

\bibitem{Ozel:2010fw}
\bibinfo{author}{\bibfnamefont{F.}~\bibnamefont{Ozel}},
  \bibinfo{author}{\bibfnamefont{G.}~\bibnamefont{Baym}}, \bibnamefont{and}
  \bibinfo{author}{\bibfnamefont{T.}~\bibnamefont{Guver}},
  \bibinfo{journal}{Phys. Rev.} \textbf{\bibinfo{volume}{D82}},
  \bibinfo{pages}{101301} (\bibinfo{year}{2010}).

\bibitem{Steiner:2010fz}
\bibinfo{author}{\bibfnamefont{A.~W.} \bibnamefont{Steiner}},
  \bibinfo{author}{\bibfnamefont{J.~M.} \bibnamefont{Lattimer}},
  \bibnamefont{and} \bibinfo{author}{\bibfnamefont{E.~F.} \bibnamefont{Brown}},
  \bibinfo{journal}{Astrophys. J.} \textbf{\bibinfo{volume}{722}},
  \bibinfo{pages}{33} (\bibinfo{year}{2010}).

\bibitem{Mueller:1996pm}
\bibinfo{author}{\bibfnamefont{H.}~\bibnamefont{Mueller}} \bibnamefont{and}
  \bibinfo{author}{\bibfnamefont{B.~D.} \bibnamefont{Serot}},
  \bibinfo{journal}{Nucl. Phys.} \textbf{\bibinfo{volume}{A606}},
  \bibinfo{pages}{508} (\bibinfo{year}{1996}).

\bibitem{Horowitz:2000xj}
\bibinfo{author}{\bibfnamefont{C.~J.} \bibnamefont{Horowitz}} \bibnamefont{and}
  \bibinfo{author}{\bibfnamefont{J.}~\bibnamefont{Piekarewicz}},
  \bibinfo{journal}{Phys. Rev. Lett.} \textbf{\bibinfo{volume}{86}},
  \bibinfo{pages}{5647} (\bibinfo{year}{2001}{\natexlab{a}}).

\bibitem{Serot:1984ey}
\bibinfo{author}{\bibfnamefont{B.~D.} \bibnamefont{Serot}} \bibnamefont{and}
  \bibinfo{author}{\bibfnamefont{J.~D.} \bibnamefont{Walecka}},
  \bibinfo{journal}{Adv. Nucl. Phys.} \textbf{\bibinfo{volume}{16}},
  \bibinfo{pages}{1} (\bibinfo{year}{1986}).

\bibitem{Serot:1997xg}
\bibinfo{author}{\bibfnamefont{B.~D.} \bibnamefont{Serot}} \bibnamefont{and}
  \bibinfo{author}{\bibfnamefont{J.~D.} \bibnamefont{Walecka}},
  \bibinfo{journal}{Int. J. Mod. Phys.} \textbf{\bibinfo{volume}{E6}},
  \bibinfo{pages}{515} (\bibinfo{year}{1997}).

\bibitem{Piekarewicz:2007dx}
\bibinfo{author}{\bibfnamefont{J.}~\bibnamefont{Piekarewicz}},
  \bibinfo{journal}{Phys. Rev.} \textbf{\bibinfo{volume}{C76}},
  \bibinfo{pages}{064310} (\bibinfo{year}{2007}).

\bibitem{Demorest:2010bx}
\bibinfo{author}{\bibfnamefont{P.}~\bibnamefont{Demorest}},
  \bibinfo{author}{\bibfnamefont{T.}~\bibnamefont{Pennucci}},
  \bibinfo{author}{\bibfnamefont{S.}~\bibnamefont{Ransom}},
  \bibinfo{author}{\bibfnamefont{M.}~\bibnamefont{Roberts}}, \bibnamefont{and}
  \bibinfo{author}{\bibfnamefont{J.}~\bibnamefont{Hessels}},
  \bibinfo{journal}{Nature} \textbf{\bibinfo{volume}{467}},
  \bibinfo{pages}{1081} (\bibinfo{year}{2010}).

\bibitem{Lattimer:2006xb}
\bibinfo{author}{\bibfnamefont{J.~M.} \bibnamefont{Lattimer}} \bibnamefont{and}
  \bibinfo{author}{\bibfnamefont{M.}~\bibnamefont{Prakash}},
  \bibinfo{journal}{Phys. Rept.} \textbf{\bibinfo{volume}{442}},
  \bibinfo{pages}{109} (\bibinfo{year}{2007}).

\bibitem{Piekarewicz:2008nh}
\bibinfo{author}{\bibfnamefont{J.}~\bibnamefont{Piekarewicz}} \bibnamefont{and}
  \bibinfo{author}{\bibfnamefont{M.}~\bibnamefont{Centelles}},
  \bibinfo{journal}{Phys. Rev.} \textbf{\bibinfo{volume}{C79}},
  \bibinfo{pages}{054311} (\bibinfo{year}{2009}).

\bibitem{Brown:2000}
\bibinfo{author}{\bibfnamefont{B.~A.} \bibnamefont{Brown}},
  \bibinfo{journal}{Phys. Rev. Lett.} \textbf{\bibinfo{volume}{85}},
  \bibinfo{pages}{5296} (\bibinfo{year}{2000}).

\bibitem{Furnstahl:2001un}
\bibinfo{author}{\bibfnamefont{R.~J.} \bibnamefont{Furnstahl}},
  \bibinfo{journal}{Nucl. Phys.} \textbf{\bibinfo{volume}{A706}},
  \bibinfo{pages}{85} (\bibinfo{year}{2002}).

\bibitem{Horowitz:2001ya}
\bibinfo{author}{\bibfnamefont{C.~J.} \bibnamefont{Horowitz}} \bibnamefont{and}
  \bibinfo{author}{\bibfnamefont{J.}~\bibnamefont{Piekarewicz}},
  \bibinfo{journal}{Phys. Rev.} \textbf{\bibinfo{volume}{C64}},
  \bibinfo{pages}{062802} (\bibinfo{year}{2001}{\natexlab{b}}).

\bibitem{RocaMaza:2011pm}
\bibinfo{author}{\bibfnamefont{X.}~\bibnamefont{Roca-Maza}},
  \bibinfo{author}{\bibfnamefont{M.}~\bibnamefont{Centelles}},
  \bibinfo{author}{\bibfnamefont{X.}~\bibnamefont{Vi\~nas}}, \bibnamefont{and}
  \bibinfo{author}{\bibfnamefont{M.}~\bibnamefont{Warda}},
  \bibinfo{journal}{Phys. Rev. Lett.} \textbf{\bibinfo{volume}{106}},
  \bibinfo{pages}{252501} (\bibinfo{year}{2011}{\natexlab{b}}).

\bibitem{Horowitz:2002mb}
\bibinfo{author}{\bibfnamefont{C.~J.} \bibnamefont{Horowitz}} \bibnamefont{and}
  \bibinfo{author}{\bibfnamefont{J.}~\bibnamefont{Piekarewicz}},
  \bibinfo{journal}{Phys. Rev.} \textbf{\bibinfo{volume}{C66}},
  \bibinfo{pages}{055803} (\bibinfo{year}{2002}).

\bibitem{Carriere:2002bx}
\bibinfo{author}{\bibfnamefont{J.}~\bibnamefont{Carriere}},
  \bibinfo{author}{\bibfnamefont{C.~J.} \bibnamefont{Horowitz}},
  \bibnamefont{and}
  \bibinfo{author}{\bibfnamefont{J.}~\bibnamefont{Piekarewicz}},
  \bibinfo{journal}{Astrophys. J.} \textbf{\bibinfo{volume}{593}},
  \bibinfo{pages}{463} (\bibinfo{year}{2003}).

\bibitem{Steiner:2004fi}
\bibinfo{author}{\bibfnamefont{A.~W.} \bibnamefont{Steiner}},
  \bibinfo{author}{\bibfnamefont{M.}~\bibnamefont{Prakash}},
  \bibinfo{author}{\bibfnamefont{J.~M.} \bibnamefont{Lattimer}},
  \bibnamefont{and} \bibinfo{author}{\bibfnamefont{P.~J.} \bibnamefont{Ellis}},
  \bibinfo{journal}{Phys. Rept.} \textbf{\bibinfo{volume}{411}},
  \bibinfo{pages}{325} (\bibinfo{year}{2005}).

\bibitem{Li:2005sr}
\bibinfo{author}{\bibfnamefont{B.-A.} \bibnamefont{Li}} \bibnamefont{and}
  \bibinfo{author}{\bibfnamefont{A.~W.} \bibnamefont{Steiner}},
  \bibinfo{journal}{Phys. Lett.} \textbf{\bibinfo{volume}{B642}},
  \bibinfo{pages}{436} (\bibinfo{year}{2006}).

\bibitem{Abrahamyan:2012gp}
\bibinfo{author}{\bibfnamefont{S.}~\bibnamefont{Abrahamyan}},
  \bibinfo{author}{\bibfnamefont{Z.}~\bibnamefont{Ahmed}},
  \bibinfo{author}{\bibfnamefont{H.}~\bibnamefont{Albataineh}},
  \bibinfo{author}{\bibfnamefont{K.}~\bibnamefont{Aniol}},
  \bibinfo{author}{\bibfnamefont{D.}~\bibnamefont{Armstrong}},
  \bibnamefont{et~al.}, \bibinfo{journal}{Phys. Rev. Lett.}
  \textbf{\bibinfo{volume}{108}}, \bibinfo{pages}{112502}
  (\bibinfo{year}{2012}).

\bibitem{Horowitz:2012tj}
\bibinfo{author}{\bibfnamefont{C.}~\bibnamefont{Horowitz}},
  \bibinfo{author}{\bibfnamefont{Z.}~\bibnamefont{Ahmed}},
  \bibinfo{author}{\bibfnamefont{C.}~\bibnamefont{Jen}},
  \bibinfo{author}{\bibfnamefont{A.}~\bibnamefont{Rakhman}},
  \bibinfo{author}{\bibfnamefont{P.}~\bibnamefont{Souder}},
  \bibnamefont{et~al.}, \bibinfo{journal}{Phys. Rev.}
  \textbf{\bibinfo{volume}{C85}}, \bibinfo{pages}{032501}
  (\bibinfo{year}{2012}).

\bibitem{Donnelly:1989qs}
\bibinfo{author}{\bibfnamefont{T.}~\bibnamefont{Donnelly}},
  \bibinfo{author}{\bibfnamefont{J.}~\bibnamefont{Dubach}}, \bibnamefont{and}
  \bibinfo{author}{\bibfnamefont{I.}~\bibnamefont{Sick}},
  \bibinfo{journal}{Nucl. Phys.} \textbf{\bibinfo{volume}{A503}},
  \bibinfo{pages}{589} (\bibinfo{year}{1989}).

\bibitem{Danielewicz:2003dd}
\bibinfo{author}{\bibfnamefont{P.}~\bibnamefont{Danielewicz}},
  \bibinfo{journal}{Nucl. Phys.} \textbf{\bibinfo{volume}{A727}},
  \bibinfo{pages}{233} (\bibinfo{year}{2003}).

\bibitem{Centelles:2008vu}
\bibinfo{author}{\bibfnamefont{M.}~\bibnamefont{Centelles}},
  \bibinfo{author}{\bibfnamefont{X.}~\bibnamefont{Roca-Maza}},
  \bibinfo{author}{\bibfnamefont{X.}~\bibnamefont{Vi\~nas}}, \bibnamefont{and}
  \bibinfo{author}{\bibfnamefont{M.}~\bibnamefont{Warda}},
  \bibinfo{journal}{Phys. Rev. Lett.} \textbf{\bibinfo{volume}{102}},
  \bibinfo{pages}{122502} (\bibinfo{year}{2009}).

\bibitem{Centelles:2010qh}
\bibinfo{author}{\bibfnamefont{M.}~\bibnamefont{Centelles}},
  \bibinfo{author}{\bibfnamefont{X.}~\bibnamefont{Roca-Maza}},
  \bibinfo{author}{\bibfnamefont{X.}~\bibnamefont{Vi\~nas}}, \bibnamefont{and}
  \bibinfo{author}{\bibfnamefont{M.}~\bibnamefont{Warda}},
  \bibinfo{journal}{Phys. Rev.} \textbf{\bibinfo{volume}{C82}},
  \bibinfo{pages}{054314} (\bibinfo{year}{2010}).

\bibitem{Pollock:1992mv}
\bibinfo{author}{\bibfnamefont{S.~J.} \bibnamefont{Pollock}},
  \bibinfo{author}{\bibfnamefont{E.~N.} \bibnamefont{Fortson}},
  \bibnamefont{and} \bibinfo{author}{\bibfnamefont{L.}~\bibnamefont{Wilets}},
  \bibinfo{journal}{Phys. Rev.} \textbf{\bibinfo{volume}{C46}},
  \bibinfo{pages}{2587} (\bibinfo{year}{1992}).

\bibitem{Sil:2005tg}
\bibinfo{author}{\bibfnamefont{T.}~\bibnamefont{Sil}},
  \bibinfo{author}{\bibfnamefont{M.}~\bibnamefont{Centelles}},
  \bibinfo{author}{\bibfnamefont{X.}~\bibnamefont{Vi\~nas}}, \bibnamefont{and}
  \bibinfo{author}{\bibfnamefont{J.}~\bibnamefont{Piekarewicz}},
  \bibinfo{journal}{Phys. Rev.} \textbf{\bibinfo{volume}{C71}},
  \bibinfo{pages}{045502} (\bibinfo{year}{2005}).

\bibitem{Tsang:2004zz}
\bibinfo{author}{\bibfnamefont{M.~B.} \bibnamefont{Tsang}}
  \bibnamefont{et~al.}, \bibinfo{journal}{Phys. Rev. Lett.}
  \textbf{\bibinfo{volume}{92}}, \bibinfo{pages}{062701}
  (\bibinfo{year}{2004}).

\bibitem{Chen:2004si}
\bibinfo{author}{\bibfnamefont{L.-W.} \bibnamefont{Chen}},
  \bibinfo{author}{\bibfnamefont{C.~M.} \bibnamefont{Ko}}, \bibnamefont{and}
  \bibinfo{author}{\bibfnamefont{B.-A.} \bibnamefont{Li}},
  \bibinfo{journal}{Phys. Rev. Lett.} \textbf{\bibinfo{volume}{94}},
  \bibinfo{pages}{032701} (\bibinfo{year}{2005}).

\bibitem{Steiner:2005rd}
\bibinfo{author}{\bibfnamefont{A.~W.} \bibnamefont{Steiner}} \bibnamefont{and}
  \bibinfo{author}{\bibfnamefont{B.-A.} \bibnamefont{Li}},
  \bibinfo{journal}{Phys. Rev.} \textbf{\bibinfo{volume}{C72}},
  \bibinfo{pages}{041601} (\bibinfo{year}{2005}).

\bibitem{Shetty:2007zg}
\bibinfo{author}{\bibfnamefont{D.~V.} \bibnamefont{Shetty}},
  \bibinfo{author}{\bibfnamefont{S.~J.} \bibnamefont{Yennello}},
  \bibnamefont{and} \bibinfo{author}{\bibfnamefont{G.~A.}
  \bibnamefont{Souliotis}}, \bibinfo{journal}{Phys. Rev.}
  \textbf{\bibinfo{volume}{C76}}, \bibinfo{pages}{024606}
  (\bibinfo{year}{2007}).

\bibitem{Tsang:2008fd}
\bibinfo{author}{\bibfnamefont{M.~B.} \bibnamefont{Tsang}}
  \bibnamefont{et~al.}, \bibinfo{journal}{Phys. Rev. Lett.}
  \textbf{\bibinfo{volume}{102}}, \bibinfo{pages}{122701}
  (\bibinfo{year}{2009}).

\bibitem{Fattoyev:2010tb}
\bibinfo{author}{\bibfnamefont{F.~J.} \bibnamefont{Fattoyev}} \bibnamefont{and}
  \bibinfo{author}{\bibfnamefont{J.}~\bibnamefont{Piekarewicz}},
  \bibinfo{journal}{Phys. Rev.} \textbf{\bibinfo{volume}{C82}},
  \bibinfo{pages}{025810} (\bibinfo{year}{2010}).

\bibitem{PREXII:2012}
\bibinfo{author}{\bibfnamefont{K.}\,\bibnamefont{Paschke}},
  \bibinfo{author}{\bibfnamefont{K.}\,\bibnamefont{Kumar}},
  \bibinfo{author}{\bibfnamefont{R.}\,\bibnamefont{Michaels}},
  \bibinfo{author}{\bibfnamefont{P.\,A.} \bibnamefont{Souder}},
  \bibnamefont{and} \bibinfo{author}{\bibfnamefont{G.\,M.}
  \bibnamefont{Urciuoli}} 
  (\bibinfo{year}{2012}),\\
  \urlprefix\url{http://hallaweb.jlab.org/parity/prex/prexII.pdf}.

\bibitem{Harakeh:2001}
\bibinfo{author}{\bibfnamefont{M.~N.} \bibnamefont{Harakeh}} \bibnamefont{and}
  \bibinfo{author}{\bibfnamefont{A.}~\bibnamefont{van~der Woude}},
  \emph{\bibinfo{title}{Giant Resonances-Fundamental High-frequency Modes of
  Nuclear Excitation}} (\bibinfo{publisher}{Clarendon, Oxford},
  \bibinfo{year}{2001}).

\bibitem{Piekarewicz:2010fa}
\bibinfo{author}{\bibfnamefont{J.}~\bibnamefont{Piekarewicz}},
  \bibinfo{journal}{Phys. Rev.} \textbf{\bibinfo{volume}{C83}},
  \bibinfo{pages}{034319} (\bibinfo{year}{2011}).

\bibitem{Piekarewicz:2012pp}
\bibinfo{author}{\bibfnamefont{J.}~\bibnamefont{Piekarewicz}},
  \bibinfo{author}{\bibfnamefont{B.}~\bibnamefont{Agrawal}},
  \bibinfo{author}{\bibfnamefont{G.}~\bibnamefont{Col\`o}},
  \bibinfo{author}{\bibfnamefont{W.}~\bibnamefont{Nazarewicz}},
  \bibinfo{author}{\bibfnamefont{N.}~\bibnamefont{Paar}}, \bibnamefont{et~al.},
  \bibinfo{journal}{Phys. Rev.} \textbf{\bibinfo{volume}{C85}},
  \bibinfo{pages}{041302(R)} (\bibinfo{year}{2012}).

\bibitem{Tamii:2011pv}
\bibinfo{author}{\bibfnamefont{A.}~\bibnamefont{Tamii}} \bibnamefont{et~al.},
  \bibinfo{journal}{Phys. Rev. Lett.} \textbf{\bibinfo{volume}{107}},
  \bibinfo{pages}{062502} (\bibinfo{year}{2011}).

\bibitem{Poltoratska:2012nf}
\bibinfo{author}{\bibfnamefont{I.}~\bibnamefont{Poltoratska}},
  \bibinfo{author}{\bibfnamefont{P.}~\bibnamefont{von Neumann-Cosel}},
  \bibinfo{author}{\bibfnamefont{A.}~\bibnamefont{Tamii}},
  \bibinfo{author}{\bibfnamefont{T.}~\bibnamefont{Adachi}},
  \bibinfo{author}{\bibfnamefont{C.}~\bibnamefont{Bertulani}},
  \bibnamefont{et~al.} (\bibinfo{year}{2012}), \eprint{\tt arXiv:1203.2155}.

\bibitem{Reinhard:2010wz}
\bibinfo{author}{\bibfnamefont{P.-G.} \bibnamefont{Reinhard}} \bibnamefont{and}
  \bibinfo{author}{\bibfnamefont{W.}~\bibnamefont{Nazarewicz}},
  \bibinfo{journal}{Phys. Rev.} \textbf{\bibinfo{volume}{C81}},
  \bibinfo{pages}{051303} (\bibinfo{year}{2010}).

\bibitem{Fattoyev:2011ns}
\bibinfo{author}{\bibfnamefont{F.}~\bibnamefont{Fattoyev}} \bibnamefont{and}
  \bibinfo{author}{\bibfnamefont{J.}~\bibnamefont{Piekarewicz}},
  \bibinfo{journal}{Phys. Rev.} \textbf{\bibinfo{volume}{C84}},
  \bibinfo{pages}{064302} (\bibinfo{year}{2011}).

\bibitem{Fattoyev:2012rm}
\bibinfo{author}{\bibfnamefont{F.}~\bibnamefont{Fattoyev}} \bibnamefont{and}
  \bibinfo{author}{\bibfnamefont{J.}~\bibnamefont{Piekarewicz}},
  \bibinfo{journal}{Phys. Rev.} \textbf{\bibinfo{volume}{C88}},
  \bibinfo{pages}{015802} (\bibinfo{year}{2012}).

\bibitem{Piekarewicz:2006ip}
\bibinfo{author}{\bibfnamefont{J.}~\bibnamefont{Piekarewicz}},
  \bibinfo{journal}{Phys. Rev.} \textbf{\bibinfo{volume}{C73}},
  \bibinfo{pages}{044325} (\bibinfo{year}{2006}).

\bibitem{Suzuki:1990}
\bibinfo{author}{\bibfnamefont{Y.}~\bibnamefont{Suzuki}},
  \bibinfo{author}{\bibfnamefont{K.}~\bibnamefont{Ikeda}}, \bibnamefont{and}
  \bibinfo{author}{\bibfnamefont{H.}~\bibnamefont{Sato}},
  \bibinfo{journal}{Prog. Theor. Phys.} \textbf{\bibinfo{volume}{83}},
  \bibinfo{pages}{180} (\bibinfo{year}{1990}).

\bibitem{VanIsacker:1992}
\bibinfo{author}{\bibfnamefont{P.}~\bibnamefont{Van~Isacker}} \bibnamefont{and}
  \bibinfo{author}{\bibfnamefont{D.~D.} \bibnamefont{Nagarajan},
  \bibfnamefont{M.~A.and~Warner}}, \bibinfo{journal}{Phys. Rev.}
  \textbf{\bibinfo{volume}{C45}}, \bibinfo{pages}{R13} (\bibinfo{year}{1992}).

\bibitem{Hamamoto:1996}
\bibinfo{author}{\bibfnamefont{I.}~\bibnamefont{Hamamoto}},
  \bibinfo{author}{\bibfnamefont{H.}~\bibnamefont{Sagawa}}, \bibnamefont{and}
  \bibinfo{author}{\bibfnamefont{X.~Z.} \bibnamefont{Zhang}},
  \bibinfo{journal}{Phys. Rev.} \textbf{\bibinfo{volume}{C53}},
  \bibinfo{pages}{765} (\bibinfo{year}{1996}).

\bibitem{Hamamoto:1998}
\bibinfo{author}{\bibfnamefont{I.}~\bibnamefont{Hamamoto}},
  \bibinfo{author}{\bibfnamefont{H.}~\bibnamefont{Sagawa}}, \bibnamefont{and}
  \bibinfo{author}{\bibfnamefont{X.~Z.} \bibnamefont{Zhang}},
  \bibinfo{journal}{Phys. Rev.} \textbf{\bibinfo{volume}{C57}},
  \bibinfo{pages}{R1064} (\bibinfo{year}{1998}).

\bibitem{Vretenar:2000yy}
\bibinfo{author}{\bibfnamefont{D.}~\bibnamefont{Vretenar}},
  \bibinfo{author}{\bibfnamefont{N.}~\bibnamefont{Paar}},
  \bibinfo{author}{\bibfnamefont{P.}~\bibnamefont{Ring}}, \bibnamefont{and}
  \bibinfo{author}{\bibfnamefont{G.~A.} \bibnamefont{Lalazissis}},
  \bibinfo{journal}{Phys. Rev.} \textbf{\bibinfo{volume}{C63}},
  \bibinfo{pages}{047301} (\bibinfo{year}{2001}{\natexlab{a}}).

\bibitem{Vretenar:2001hs}
\bibinfo{author}{\bibfnamefont{D.}~\bibnamefont{Vretenar}},
  \bibinfo{author}{\bibfnamefont{N.}~\bibnamefont{Paar}},
  \bibinfo{author}{\bibfnamefont{P.}~\bibnamefont{Ring}}, \bibnamefont{and}
  \bibinfo{author}{\bibfnamefont{G.~A.} \bibnamefont{Lalazissis}},
  \bibinfo{journal}{Nucl. Phys.} \textbf{\bibinfo{volume}{A692}},
  \bibinfo{pages}{496} (\bibinfo{year}{2001}{\natexlab{b}}).

\bibitem{Paar:2004gr}
\bibinfo{author}{\bibfnamefont{N.}~\bibnamefont{Paar}},
  \bibinfo{author}{\bibfnamefont{T.}~\bibnamefont{Niksic}},
  \bibinfo{author}{\bibfnamefont{D.}~\bibnamefont{Vretenar}}, \bibnamefont{and}
  \bibinfo{author}{\bibfnamefont{P.}~\bibnamefont{Ring}},
  \bibinfo{journal}{Phys. Lett.} \textbf{\bibinfo{volume}{B606}},
  \bibinfo{pages}{288} (\bibinfo{year}{2005}).

\bibitem{Tsoneva:2003gv}
\bibinfo{author}{\bibfnamefont{N.}~\bibnamefont{Tsoneva}},
  \bibinfo{author}{\bibfnamefont{H.}~\bibnamefont{Lenske}}, \bibnamefont{and}
  \bibinfo{author}{\bibfnamefont{C.}~\bibnamefont{Stoyanov}},
  \bibinfo{journal}{Phys. Lett.} \textbf{\bibinfo{volume}{B586}},
  \bibinfo{pages}{213} (\bibinfo{year}{2004}).

\bibitem{Tsoneva:2007fk}
\bibinfo{author}{\bibfnamefont{N.}~\bibnamefont{Tsoneva}} \bibnamefont{and}
  \bibinfo{author}{\bibfnamefont{H.}~\bibnamefont{Lenske}},
  \bibinfo{journal}{Phys. Rev.} \textbf{\bibinfo{volume}{C77}},
  \bibinfo{pages}{024321} (\bibinfo{year}{2008}).

\bibitem{Klimkiewicz:2007zz}
\bibinfo{author}{\bibfnamefont{A.}~\bibnamefont{Klimkiewicz}}
  \bibnamefont{et~al.}, \bibinfo{journal}{Phys. Rev.}
  \textbf{\bibinfo{volume}{C76}}, \bibinfo{pages}{051603}
  (\bibinfo{year}{2007}).

\bibitem{Carbone:2010az}
\bibinfo{author}{\bibfnamefont{A.}~\bibnamefont{Carbone}},
  \bibinfo{author}{\bibfnamefont{G.}~\bibnamefont{Colo}},
  \bibinfo{author}{\bibfnamefont{A.}~\bibnamefont{Bracco}},
  \bibinfo{author}{\bibfnamefont{L.-G.} \bibnamefont{Cao}},
  \bibinfo{author}{\bibfnamefont{P.~F.} \bibnamefont{Bortignon}},
  \bibnamefont{et~al.}, \bibinfo{journal}{Phys. Rev.}
  \textbf{\bibinfo{volume}{C81}}, \bibinfo{pages}{041301}
  (\bibinfo{year}{2010}).

\bibitem{Adrich:2005}
\bibinfo{author}{\bibfnamefont{P.}~\bibnamefont{Adrich}} \bibnamefont{et~al.},
  \bibinfo{journal}{Phys. Rev. Lett} \textbf{\bibinfo{volume}{95}},
  \bibinfo{pages}{132501} (\bibinfo{year}{2005}).

\bibitem{Wieland:2009}
\bibinfo{author}{\bibfnamefont{O.}~\bibnamefont{Wieland}} \bibnamefont{et~al.},
  \bibinfo{journal}{Phys. Rev. Lett.} \textbf{\bibinfo{volume}{102}},
  \bibinfo{pages}{092502} (\bibinfo{year}{2009}).

\bibitem{Bertsch:1994}
\bibinfo{author}{\bibfnamefont{G.~F.} \bibnamefont{Bertsch}} \bibnamefont{and}
  \bibinfo{author}{\bibfnamefont{R.~A.} \bibnamefont{Broglia}},
  \emph{\bibinfo{title}{Oscillations in Finite Quantum Systems}}
  (\bibinfo{publisher}{Cambridge University Press, Cambridge},
  \bibinfo{year}{1994}).

\bibitem{Paar:2007bk}
\bibinfo{author}{\bibfnamefont{N.}~\bibnamefont{Paar}},
  \bibinfo{author}{\bibfnamefont{D.}~\bibnamefont{Vretenar}},
  \bibinfo{author}{\bibfnamefont{E.}~\bibnamefont{Khan}}, \bibnamefont{and}
  \bibinfo{author}{\bibfnamefont{G.}~\bibnamefont{Colo}},
  \bibinfo{journal}{Rept. Prog. Phys.} \textbf{\bibinfo{volume}{70}},
  \bibinfo{pages}{691} (\bibinfo{year}{2007}).

\bibitem{Paar:2010ww}
\bibinfo{author}{\bibfnamefont{N.}~\bibnamefont{Paar}}, \bibinfo{journal}{J.
  Phys.} \textbf{\bibinfo{volume}{G37}}, \bibinfo{pages}{064014}
  (\bibinfo{year}{2010}).

\end{thebibliography}
\end{document}